\documentstyle[12pt,aasms4]{article}

\begin{document}

\title{PRODUCTION OF LITHIUM IN THE GALACTIC DISK.}

\author{E. Casuso$^{1}$}
\and
\author{J. E. Beckman$^{1,2}$}.

\affil{$^{1}$Instituto de Astrofisica de Canarias,E-38200 La Laguna,
Tenerife, Spain.E-mail: eca@ll.iac.es}

\affil{$^{2}$Consejo Superior de Investigaciones Cientificas.CSIC.Spain}
\affil{TO APPEAR IN PASP JULY 2000} 
\journalid{Vol}{Journ. Date}
\articleid{start page}{end page}
\paperid{manuscript id}
\cpright{type}{year}
\ccc{code}
\lefthead{Casuso & Beckman}
\righthead{CHEMICAL EVOLUTION}

\begin{abstract}

The abundance of Li in stars formed within the past 5 Gyr is logN(Li)
= 3.2($\pm$0.2),while the corresponding value for the oldest stars in
the Galaxy is logN(Li) = 2.2($\pm$0.2).The global evidence suggests
that the latter represents the full, or the major part of the
primordial abundance, so that the difference of an order of magnitude
is due to Li produced in the Galaxy.  It is well known that spallation
of insterstellar CNO by $^{4}$He and protons in galactic cosmic rays
(GCR) can produce Li,but models yield a shortfall of almost an order
of magnitude compared with the current observed abundance
range.Another GCR reaction, $\alpha$+$\alpha$ fusion has been invoked
to explain some Li production in the early Galaxy,but application of
this to the disk yielded too much early Li or too little current
Li.These failures led to a search for alternative
mechanisms,essentially stellar, at particular phases of evolution:the
helium flash phase in AGB stars,in novae,and during supernova.  Here
we stress the importance of the observed upper envelope in the plot of
Li v. Fe in stars as a constraint on any mechanism in any model aiming
to account for disk Li.  We show that a good can be found assuming
that low energy GCRs produce the Li,with the $\alpha$+$\alpha$
reaction as the key mechanism although production in supernovae cannot
at this stage be excluded.There is an apparent time delay in the Li
production, relative to O and Fe, which if confirmed could be
explained by the origin of a low energy $\alpha$-particle component in
processes associated with stars of intermediate and low mass.The
$\alpha$ flux at a given epoch would then be proportional to the
amount of gas expelled by low and intermediate mass stars in the
Galaxy, though the acceleration of these alphas could still be linked
to more energetic events as SN explosions.  The present scenario
appears to account coherently for the closely related observations of
the temporal evolution in the Galaxy (Halo+Disk) of abundances of
$^{12}$C,$^{13}$C,$^{14}$N,$^{16}$O,$^{26}$Fe, the two main peaks (one
in the Halo and one in the Disk) in the G-dwarf stellar frequency
distribution, and the evolution of $^{9}$Be and $^{10}$B+$^{11}$B via
GCR spallation reactions without requiring the very high local
cosmic-ray fluxes implied by the spallation close to SN (Casuso $\&$
Beckman 1997).Adding a natural mechanism of differential depletion in
red supergiant envelopes, we can explain the observed time evolution
of the abundance of D and that of the isotopic ratios
$^{7}$Li/$^{6}$Li and $^{11}$B/$^{10}$B (Casuso $\&$ Beckman 1999)
starting from an SBBN model with baryon density $\sim$0.05.  Our model
also predicts the second Li-"plateau" found for [Fe/H] between -0.2
and +0.2, due to the "loop back" implied for Li (also for $^{9}$Be and
B) because of the required infall of low metallicity gas to the disk.
Without ruling out other mechanisms for the main production of Li in
the Galactic Disk, the low-energy $\alpha$+$\alpha$ fusion reaction in
the ISM offers a promising contribution.

\end{abstract}


\section{Introduction and observational base}

A few seconds after the Big Bang, four light isotopes were produced:
D,$^{3}$He,$^{4}$He and $^{7}$Li (see eg. Walker et al. 1991,Copi et
al. 1995,Shramm and Turner 1998); all of these warrant careful study,
and here we are focusing on $^{7}$Li.The importance of understanding
the evolution of the Galactic abundance of Li was highlighted in the
key discovery by Spite $\&$ Spite (SS) (1982) that the observed
abundance of Li in Galactic stars does not continue to fall uniformily
with decreasing iron abundance below [Fe/H]$\simeq$-1, but levels off
to a "plateau" at a level of logN($^{7}$Li)$\simeq$2, which SS
interpreted as corresponding to the abundance produced by big bang
nucleosynthesis (SBBN).Spite and Spite measured the $^{7}$Li abundance
as a function of metallicity (iron abundance) and surface
temperature.They found that the $^{7}$Li abundance is flat for surface
temperatures greater than about 5600K, and further, it is also flat
for the stars with the lowest iron abundance.The first plateau
suggests that the stars with the highest surface temperatures are not
destroying their $^{7}$Li by convection (the depth of the convective
zone depends on surface temperature and is shallowest for stars with
the highest surface temperatures).The second plateau indicates that
any post-big-bang production must be insignificant for the most
metal-poor stars because the $^{7}$Li abundance does not increase with
iron abundance.The case against major depletion (and hence for a
plateau abundance that reflects the primeval abundance) was
strengthened by the observation of $^{6}$Li in certain population II
stars (Smith et al. 1993, Hobbs and Thorburn 1994).Big-bang production
of $^{6}$Li is negligible; the $^{6}$Li seen was probably produced by
cosmic-ray processes (along with beryllium and boron).Because $^{6}$Li
is much more fragile than $^{7}$Li and yet still survived with the
abundance relative to Be and B expected from cosmic-ray production,
depletion of of $^{7}$Li cannot have been very significant (Steigman
et al. 1993).  Using this interpretation the primordial abundance is
given by logN$_{P}$($^{7}$Li)=2.2($\pm$0.2), a value confirmed in
detailed work by a succession of authors (Rebolo,Beckman $\&$ Molaro
1987,Hobbs $\&$ Thorburn 1991,Spite 1991,Thorburn 1994) which can be
combined with the SBBN produced abundance of $^{4}$He (see e.g. Pagel
et al. 1992), to infer basic cosmological parameters: the universal
baryon density $\Omega$$_{b}$, and the number of massless
two-component neutrino types N$_{\nu}$.To be sure that the population
II abundance of $^{7}$Li is a largely undepleted SBBN abundance,
entails two essential steps: showing that population II stars (with
T$_{eff}$$\geq$5500 K) have not depleted or barely depleted their
$^{7}$Li, and showing that most of the $^{7}$Li in population I stars
is of Galactic origin.The first step has already been accomplished via
the theoretical work of the Yale group, who showed (Pinsonneault,
Deliyannis $\&$ Demarque 1992) that sub-surface convective transport,
and hence $^{7}$Li depletion is strongly suppressed at low
metallicities.As a result of this, and of steadily accumulating
observations, opinion (see Spite $\&$ Spite 1993) has swung strongly
behind the view that logN($^{7}$Li)$\simeq$2.2 is the SBBN
value.Thorburn's (1994) refined work on the "plateau", has brought out
a scatter in the $^{7}$Li v. [Fe/H] plot below [Fe/H]=-1.5, which is
incompatible with zero production of $^{7}$Li in the halo, but in
practice strongly supports a primordial value for $^{7}$Li not far
above logN$_{P}$($^{7}$Li)$\simeq$2.  The second step is quite
complicated, because during the disk lifetime there may have been a
number of significant production processes for Li, and also a number
of destruction, or depletion processes.The initial primordial
abundance masks any Li evolution in the halo, so we concentrate our
attention in the present paper on production in the disk and its
interpretation.  The evolution of the lithium abundance in the
Galactic disk can be followed via observations which define the upper
envelope of the lithium abundance in stars over the range of iron
metallicity, -1.5${\leq}$[Fe/H]${\leq}$0.1, which characterizes the
disk population.The underlying assumption is that while lithium is in
general depleted within stars, for a given value of metallicity the
highest observed abundance value for a set of stars will correspond to
minimum depletion, and hence to an optimum approximation to the
Galactic interstellar lithium abundance at the epoch when the stars
were formed.Following the evolution of lithium should give similar
insight into the processes which form it to that which we can obtain
by following the evolution of any other element.The rise in the ratio
O/Fe with decreasing Fe, for example, gives the key to understanding
the origin of a major fraction of Galactic oxygen in supernovae of
type II, whereas iron is formed in all stars with masses greater than
or equal to 1 solar mass.

Although the general trend in Galactic Li evolution can be followed via the Li-Fe envelope, the effect of depletion imposes the
need for the  greatest care when interpreting the observed abundance in
any single object.Depletion is well-known to occur in cool stars (see
e.g. Pinsonneault et al. 1992,Deliyannis et al. 1990): those with
T$_{eff}$ less than the solar value, and occurs also in the "{\it
lithium gap}" (Cayrel et al. (1984), Boesgaard (1987)) in the middle-F
range of spectral classes.Stellar depletion renders lithium
particularly interesting as a probe of stellar structure (Steigman et al. 1993),but makes it
more difficult to interpret measured abundances in terms of production
processes.

However, if we are not able to account adequately for the present-day
and post-solar system abundance values: logN(Li)$\geq$3, there must
remain some room for doubt about the BBN value.For this reason, as
well as for its intrinsic importance as a test of Galactic evolution
models, the source(s) of Galactic lithium continue to be of
considerable research interest.A number of production mechanisms have
been proposed: cosmic ray spallation of CNO (Reeves,Fowler \& Hoyle
(1970), Meneguzzi,Audouze \& Reeves (1971), Walther,Mathews \& Viola
(1989)), nucleosynthesis in novae (Arnould \& Norgaard (1979),
Starrfield et al. (1978)), in the atmospheres of red giants (Cameron
\& Fowler (1971)), in supernovae (Dearborn et al.  1989,Woosley et
al. 1990), and in AGB stars and carbon stars (D'Antona and Matteucci
(1991)) and in black hole binaries (Martin et al.  (1994)).It is well
accepted that processes in the interior of normal stars not only fail
to yield lithium, but tend to deplete it.In spite of the detection of
individual lithium rich objects which might be characteristic sources,
models which incorporate such sources into a Galactic evolution scheme
(Audouze et al. (1983), Abia \& Canal (1988), D'Antona \& Matteucci
(1991)) do not give good agreement with the observed lithium-iron
envelope.Further, the spatial homogeneity of the Fe-Li curve points
against sparse sets of point sources, even distributed sources, and in
favour of a more diffuse origin for the lithium.  Recent detailed
models of Li production assumed in carbon stars, massive AGB
stars,SNII and novae (Romano 1999), do not give fair fits to the very
high slope of the Li abundance vs. [Fe/H] near [Fe/H]$\simeq$-0.3 (see
Fig. 1 of Romano 1999).

The light nuclide $^{6}$Li is not produced dignificantly in SBBN and
is expected to be produced over the lifetime of the Galaxy in Galactic
cosmic ray spallation as well as $\alpha$+$\alpha$ fusion
reactions.Its high fragility to stellar processing makes it a less
useful tool than $^{7}$Li to constrain big bang nucleosynthesis, but
many authors have modelled $^{6}$Li time evolution due to the assumed
conexion with the $^{9}$Be and B abundances (Yoshii et
al. 1997,Lemoine et al. 1997,Vangioni-Flam et al. 1999,Fields and
Olive 1999, Ryan et al. 1999).

It has been suggested (see,e.g.,Steigman et al. (1993)) that since
(due to dust grain depletion, and ionization equilibrium uncertainly)
isotope ratios can be determined more reliably in the interstellar
medium than absolute abundances or ratios of different elements, the
interstellar isotope ratio $^{6}$Li/$^{7}$Li might offer a better
parameter to test source models than the absolute lithium abundance
estimated directly in the ISM.However, measurements of the local
interstellar $^{7}$Li/$^{6}$Li ratio (e.g. Lemoine et al.
(1993),Meyer,Hawkins \&
Wright (1993)) show major variations, with differences of up to an
order of magnitude from one IS cloud to another.Further, given the
extreme difficulty of the ratio measurement in a stellar atmosphere and
the consequent extreme paucity of such data as a function of
metallicity together with the difficult interpretation of these data in
terms of differential stellar depletion as a function of stellar
surface temperature, it would be especially risky to attempt to draw
conclusions at this stage by using a chemical evolution model to
predict the evolution of the isotope ratio against, say, iron
abundance.Because of the apparent
spatial inhomogeneity it is not even safe to place too much emphasis on
the well measured solar system $^{7}$Li/$^{6}$Li ratio of 12.5 (Mason
(1971)).These considerations have led us to the present approach of
concentrating on the overall Li abundance envelope as a key model
constraint.

In this paper we adopt the technique we were the first to use in Rebolo
et al.  (1988) of assuming that the upper envelope of lithium vs. iron
abundance plot is at least a close approximation to the undepleted
curve.We do in fact examine the alternative hypothesis: that this
envelope represents a depletion curve, and show that this
interpretation is quantitatively improbable, so leaving the way clear
for the use of the lithium vs. iron envelope as a test of lithium
production processes.The purpose of the paper is to show, using this
envelope,
 which types of production processes are excluded, and which
 permitted.Without going into any numerical detail it is evident from
 inspection (see Fig. 1) that the rise in the lithium abundance towards
 the values found in objects close to solar (iron) metallicity occurs relatively
 late in the Galactic disk evolution time scale; the lithium rise lags the
 rise in iron, precisely the opposite case to that of oxygen (see Fig.
 3).A direct implication is that processes associated with type II
 supernovae could, but with difficulty, yield the observed lithium production.This consideration not only
 covers hypothetical processes within the supernovae, but interactions
 of the energetic particles which they produce in processes occurring
 in the
interstellar medium (ISM).This is just an example of
how we can hope to constrain the Galactic lithium production process using the
available observational data.Below we will use quantitative modelling
(both analytical and numerical) with the aim of reproducing the Li-Fe
envelope, thereby eliminating processes which predict significantly
different envelopes.What remains will be candidate material for the
process (or processes) which gave rise to some 90$\%$ of the
lithium we can observe today.

In section 2 we show, using simplified analytical models, how the
overall shape of the lithium-iron curve for the Galactic disk can be
reproduced on the assumptions of delayed production of lithium and
increasing infall of gas to the disk.In section 3 we describe briefly
a numerical chemical evolution model used to handle the detailed
evolution of lithium.In section 4 we examine quantitatively the problem
of the galactic cosmic ray (GCR) flux as a candidate source for
lithium.In section 5 we compare some of the suggested production
mechanisms for lithium.Finally we
draw some conclusions about the primordial abundance of lithium.

\section{Analytical and semi-analytical models for the temporal
evolution of Li: comparison with data and with previous modelling.}

Numerical modelling of Galactic chemical evolution, which we consider
in section 3, offers the advantage of being able in principle to match
with realism the physical variables which have driven it, in other
words given correct assumptions to give exact fits to the relevant
observations.However, without descriptions of inordinate length and
including full listing of complete codes, such numerical models are
not as transparent as desired.While analytical models are inevitably
too simple to reproduce most data sets, their use is more didactic,
enabling the underlying physics to be better demonstrated.For this
reason we have chosen first to show analytical models which illustrate
the physical requirements of any scheme that predicts the observed
lithium-iron relation, before presenting our numerical models.The
analytical models are designed according to standard methodology,
which is based on the paradigmatic work of Tinsley (1980).  As the
observations give us directly the evolution of the abundance of one
element vs. that of another, and because the analytical treatment
predicts the evolution of abundances vs. time, we adopt hereafter our
well tested (see Figs. 2a,2b) numerical results for the translation
from the metallicity (taken as [Fe/H] or [O/H]) plane to the time
plane and vice versa.  Firstly, in the volume under consideration, we
set the star formation rate, SFR, proportional to the gas fraction
${\sigma}_{g}$, following Schmidt (1959), so that

\begin{equation}
SFR(t)={\gamma}{\cdot}{\sigma}{_{g}}(t)
\end{equation}
and the time evolution of the star formation rate is given by
\begin{equation}
{\frac{dSFR(t)}{dt}}=-{\gamma}{\cdot}SFR(t)+{\gamma}{\cdot}E(t)
\end{equation}
where E(t) is the gas acquired by the zone per unit time due to
expulsion by stars plus any net infall of gas to the volume , and
SFR(t) is the rate of conversion of gaseous mass into stellar
mass.Integrating equation (2) gives

\begin{equation}
SFR(t)={\gamma}{\cdot}e{^{{\gamma}(-t+{\int}G(t)dt)}}
\end{equation}
where
\begin{equation}
G(t)=\frac{E(t)}{SFR(t)}
\end{equation}
To simplify the treatment we will first approximate E(t) to SFR(t);
this is the case, for example, where the star forming process has
efficiency close to 100$\%$ and rapidly consumes all the available gas
in the volume so that in each time interval new star formation uses
only gas expelled from existing stars.This is akin to the assumption of
instantaneous recycling.In this case G(t)$\sim$1 and
SFR(t)$\sim$$\gamma$.A similar expression would be obtained if the
infalling
mass of gas added to the gas expulsion by stars, at time t, were
comparable to the gas consumption by star formation at the same time
t.The first assumption to test is that lithium is produced either in
SNe of type II or by processes in the interstellar medium caused by
energetic particles expelled from type II supernovae.In this case
the rate of lithium production will be proportional to the star
formation rate, which gives

\begin{equation}
{\frac{dLi(t)}{dt}}{\propto}SFR(t){\propto}{\gamma}
\end{equation}
which integrating and translating from the time plane to the
metallicity plane through simple parabolic fit (of the form
[Fe/H]$\simeq$-(1-$\frac{t}{15}$)$^{2}$ with t in Gyr.) to the data of
Fig.2a, gives
\begin{equation}
Li(t){\simeq}Li(0)+{\gamma}(1-(-[Fe/H]){^{1/2}})
\end{equation}
where Li(0) is the initial lithium abundance.From this we can deduce that models
in which the bulk of Galactic Li is produced in processes involving
type II SNe could satisfy the
observational requirements, but with some degree of difficulty (see Fig.3).

One can obtain an approximation to the effects of infall in this type
of models in a less direct way, but which serves to illustrate the
principle.Here we must refer ahead to a numerical model developed in
Section 3, from which we take an approximate time-dependence of the
rate of SNeII.It turns out to be approximately parabolic (see Fig. 5)
centered at t=80 in unit model steps of 100 Myr, and we fit

\begin{equation}
\frac{dLi(t)}{dt}{\propto}1.5{\cdot}10{^{-8}}(t-80){^{2}}+5{\cdot}10{^{-5}}
\end{equation}

which gives on integration

\begin{equation}
Li(t){\propto}0.5{\cdot}10{^{-8}}(t-80){^{3}}+5{\cdot}10{^{-5}}t+Li(0)
\end{equation}

The numerical coefficients show that for all times of interest (t$\geq$5Gyr) up to the present
age of the disk,the cubic term can in fact be neglected, and the
lithium abundance grows essentially as

\begin{equation}
Li(t){\propto}5{\cdot}10{^{-5}}t+Li(0)
\end{equation}
which shows the same behaviour than that of eq. (6) in the metallicity plane.

Thus processes whose rate is proportional to the number of SNII in the
disk (either in stars or in the ISM) are not ruled out by the lithium-iron
envelope test, in all scenarios, with or without infall of gas.

Processes which depend on the SNII rate included among current explanations for lithium
production in the ISM (see e.g. Ramaty et al. 1997): spallation of CNO by highly energetic alphas, has been adduced to
account for the major fraction of the lithium produced.An
alternative non-stellar mechanism is the interaction of moderate energy
alphas with the
existing abundant He nuclei in the ISM, producing lithium via the
$\alpha$+$\alpha$ fusion reaction.The sources of these low
energy alphas can be the winds of normal stars (see section 4).In this case an additional IS acceleration mechanism is
required to give the alphas sufficient energy, at least a few MeV,
required for the $\alpha$+$\alpha$ fusion reaction.One could assume that the effects of the
presence of supernovae on the ISM can produce the required acceleration
(but, we have shown, using simple calculations, that the same
results would follow using the acceleration in wind termination shocks
of stars of all masses as proposed by Rosner $\&$ Bodo (1996)).The
lithium production rate at a given epoch will then be proportional to
the mass outflow from stellar winds multiplied by the supernova rate.We
can approximate the latter as constant (cf. above) and the mass outflow
rate from stars of a given mass will be proportional to the number of
stars of that mass; for low mass stars (those with masses less than
1M$_{\odot}$, and hence supplying gas only via winds because their
lifetimes are greater than the life of the disk) this number is fully
cumulative, and we have approximately (integrating in time
SFR(t)=constant)

\begin{equation}
N{_{stars}}(t){\propto}t
\end{equation}

and so

\begin{equation}
\frac{dLi(t)}{dt}{\propto}t
\end{equation}

hence, integrating, and translating from the t-plane to the metallicity plane through the same fit to data than was taken for eq. (6), one has:

\begin{equation}
Li(t){\simeq}Li(0)+{\gamma}'(1-(-[Fe/H]){^{1/2}}){^{2}}
\end{equation}
with $\gamma$' a constant;
see Fig. 3 to compare the predictions of eq. (12) with data.
Now, considering intermediate mass stars (those with masses
1M$_{\odot}$$\leq$m$\leq$3M$_{\odot}$, which supply gas mainly via
processes in their late evolutionary stages) as the main producers of
Li, because of their fairly long lifetimes, their expulsion of gas accumulates
over times long after their birth, and for the lower part of this mass
range, keeps accumulating until the present epoch.We may approximate analytically the collective gas expulsion rate by an
exponential, while SN(t) can be taken
as approximately constant, as above.In this case one has

\begin{equation}
\frac{dLi(t)}{dt}{\propto}e{^{kt}}
\end{equation}

and integrating and translating to the metallicity plane as above, we have

\begin{equation}
Li(t){\simeq}{\gamma}''(1+e{^{k'(1-(-[Fe/H]){^{1/2}})}})
\end{equation}
with $\gamma$'' and k' constants.

In Fig. 3 we can see the fit of this expression to the observations.
One can go further and analyze in more detail the time dependence of
the Li abundance on the properties of stellar mass ranges as follows:
Assuming an SFR approximately constant, and assuming that the
produccion of Li is associated with the gas expulsion by stars of all
masses one has

\begin{equation}       
\frac{d(Li(t))}{dt}{\propto}{\int_{m_{t}}^{m_{u}}}m{^{-2.35}}R(m)dm
\end{equation} 

Taking an analytical approximation for R(m) in the form m$^{0.2}$-0.58
(based on the numerical values of Renzini and Voli (1981)) and
integrating, one has

\begin{equation}
\frac{d(Li(t))}{dt}{\propto}-0.005+{\frac{m{_{t}}{^{-1.15}}}{1.15}}-{\frac{0.58}{1.35}}m{_{t}}{^{-1.35}}
\end{equation}

Using the approximation relating the mass of a star and its lifetime as
m${_{t}}{\simeq}(11.7){^{1/2}}t{^{-1/2}}$, translating to the metallicity plane as below, and integrating one has, neglecting the
first term which has a very low value compared with the other two:

\begin{equation}
Li(t){\simeq}Li(0)+k''(-0.005(1-(-[Fe/H]){^{1/2}})-0.049(1-(-[Fe/H]){^{1/2}}){^{1.675}}+0.13(1-(-[Fe/H]){^{1/2}}){^{1.575}})
\end{equation}
where k'' is a constant;
see Fig. 3.If we place an upper
limit on the mass range of stars contributing to the gas which yields
Li, this analytical approximation leads to a sharp increase from zero
Li production on short timescales to a high value when times reach the
scale of lifetime of the stars
whose masses are those of the upper limit.For example for an upper mass
limit of 1M$_{\odot}$ the Li production will be zero until the time is
near 12Gyr or [Fe/H]$\simeq$0.0 (see Fig. 3).Thus we can see, in general terms, that using a
gas expulsion timescale and modulating the upper mass limit, we can
find solutions leading to late-time production of Li, as apparently required by
the observations.

For comparison one can set out a similar formulation using the
assumption that Li production is proportional to the cumulative number
of stars of all masses at a given time, which implies production within the stars or
in their envelopes, rather than in the ISM via expelled gas:

\begin{equation}       
\frac{d(Li(t))}{dt}){\propto}{\int_{m_{L}}^{m_{t}}}m{^{-2.35}}dm
\end{equation}

Performing these integrals with the same approximation for the mass-time and time metallicity
dependences as before, and taking m$_{L}$=0.1M$_{\odot}$, one has

\begin{equation}
Li(t){\simeq}Li(0)+k'''(-0.084(1-(-[Fe/H]){^{1/2}}){^{1.675}}+16.58(1-(-[Fe/H]){^{1/2}}))
\end{equation}
where k''' is a constant.
A comparison of these stylized models: with Li production proportional
to the cumulative numbers of moderate mass stars or alternatively to the cumulative flux
of expelled gas, is given in Fig. 3.

In all the approximations based on proportionality of Li production
rate to the gas expulsion rate by stars of low or intermediate masses
(equations (12), (14)) one can see considerable resemblance
to the observed lithium growth profile.The reason for the importance of intermediate mass stars, rather
than high mass stars, as producers of alpha particles leading to Li
production is that the former turn out to be particularly efficient in
expelling He. For stars with masses greater than 3M$_{\odot}$ the
central temperature becomes high enough for the ignition of the
triple-alpha reaction (which transforms He to C) before the giant stage
is reached.At the moderate densities of the central regions, this
nuclear process gains importance in a gradual manner.On the other hand
for stars with masses between 0.5M$_{\odot}$ and $\sim$3M$_{\odot}$ the
central regions become degenerate and the triple-alpha reaction ignites
via the violent {\it helium flash}.

Although we would certainly not wish at this stage to exclude the SNe
as key producers of Li in the Galactic Disk, to be coherent with our
general scenario of chemical evolution for the Galaxy, we have used
the $\alpha$+$\alpha$ process in the ISM as illustrative of processes
which follow the behaviour of stars of intermediate and low masses,
those whose lifetimes are long, and whose numbers in the disk have
therefore grown cumulatively with time.Any process with equivalent
time-dependent characteristics might, in general terms, satisfy this
global observational constraint.Production of Li in novae (Arnould
$\&$ Norgaard 1979,Starrfield et al. 1978), or in SN type I, might
also satisfy the criterion of delayed production because there the
rate of production would be proportional to the number of stars with
intermediate masses at each time (cumulative as their lifetimes are
long).But, as one can see in Fig. 4, where we plot the results from
different numerical models in the case of GCR flux proportional to the
cumulative number of stars with masses less than or equal to
1M$_{\odot}$ (for upper mass limits greater than 1M$_{\odot}$ the
increase in the number of stars with time is proportionally less and
less), the prediction, although not bad, is clearly inferior in fit to
that obtained using gas expulsion by stars of masses less than or
equal to 3M$_{\odot}$ (and is in fact also worse than that obtained
using gas expulsion of stars with masses less than or equal to
2M$_{\odot}$).

Another possibility which has been discussed is the production of Li in
compact objects such as neutron stars and black holes, but the time
evolution of the numbers of compact objects would be proportional to
the rate of gas expulsion by all stars (which is mainly that of
SNI+SNII).This production is shown in Fig. 1 from our numerical
modelling.In fact, as one would expect,the curve  is
similar to that for the model in which Li production is proportional to
the SFR, also shown in Fig. 1.

Further possibilities for the production of Li, such as those in AGB
stars or carbon stars (Matteucci et al. (1995)) are unable to
satisfy the detailed observational constraints, as will be seen in Fig.
5.

The aim of this section has been to show those categories of models,
and hence of lithium sources, which can best account
for the lithium-iron envelope observations.However for a valid test of any
process which is a candidate
to have produced the observed disk lithium we have no choice but to use
quantitative, numerical, modelling methods.

\section{Numerical evolutionary models: the basic formalism.} 

The model we have used for the evolution of the disk in the solar
neighborhood empoys the formalism already explained in Casuso \&
Beckman (1997),which embodies a numerical rather than an analytical
approach,in order to take all the relevant physics adequately into account.The model allows us to follow the
evolution,within a fixed volume of space,of the gaseous mass fraction
$\sigma_g$ and the abundances X$_i$ of 6
nuclides:$^{4}$He,$^{12}$C,$^{13}$C,$^{14}$N,$^{16}$O, and
$^{56}$Fe,selected because observations of their evolutionary abundance
behaviour are available.The set of basic equations employed,in which
the units are mass fraction per unit time interval,are:

\begin{equation}
        d{\sigma_g} = -SFR(t) + E(t)                                     
\end{equation}
\begin{equation}        d({\sigma_g}X{_i})={\int_{m_{t}}^{m_{u}}}SFR(t-t{_m}){\phi}(m)(Q{_i}(m)+X{_i}(t-t{_m})(R(m)-Q{_i}(m))-R(m)X{_i}(t))dm+J(t)
\end{equation}                                                      
with 
\begin{equation} 
J(t) = P(t)(X{_i}'(t)-X{_i}(t))
\end{equation}
with  
\begin{equation}  
E(t) = {\int_{m_{t}}^{m_{u}}}SFR(t-t{_m}){\phi}(m)R(m)dm + P(t)                
\end{equation}
in which t$_m$ is the lifetime of a star of mass m,X$_i$'(t) is the
halo abundance,P(t) is a term which represents net inflow of material
to the volume under study, SFR(t) is the star formation rate,and
$\phi$(m) the initial mass function (IMF) of the stars.Within this
volume there is a population of stars whose lifetimes t$_m$,for a mass
m$\geq$m$_t$ (where m$_t$ is the mass of a star which has a lifetime
t),are less than or equal to the value of the time variable t,and which
eject the products of their internal nucleosynthesis at a rate
proportional to SFR(t-t$_m$),the star formation rate at their birth.We
term the
 fraction of its mass which a star ejects during its lifetime R(m),and
 Q$_i$(m) is the yield of nuclide i from a star of mass m.The total
 mass which has been added to the ISM,either by stellar evolution or by
 net inflow into the volume under consideration,is called E(t).All
 relevant stages of stellar evolution have been taken into account,
 including post-main-sequence phases (e.g. AGB-stars, planetary nebulae
 and other gas ejection stages) and explosive processes.

We have used a simple proportionality law for the dependence of the
star formation rate on the gas fraction,viz.
SFR(t)=$\gamma\sigma_g$$^k$(t) where,following the classical approach
of Schmidt (1959) we have used k=1,and the value of $\gamma$ is 0.11
Gyr$^{-1}$.The observed parameters of chemical evolution for the solar
neighborhood are reasonably reproduced.As a suitable approximation to
the IMF we have used the Salpeter (1955) law,i.e. $\phi$(m)
$\propto$ m$^{-(1+x)}$ with x=1.35,between 73 M$_\odot$ and 0.5
M$_\odot$,and have approximated the flattening observed at low masses
(see e.g. Scalo (1986),Kroupa et al. (1993)) with a plateau of value
$\phi$(0.5) between 0.5M$_\odot$ and 0.1M$_\odot$.This approximation is
convenient for computations,and represents a reasonable fit to the
observations.Slightly better but more complicated fits to the
observations would not affect any of the conclusions reached
here.Finally we approximated the stellar lifetime t$_m$ as a function
of mass m,following Arimoto $\&$ Yoshii (1986) by t$_m$=11700/m$^{2}$
in units of Myrs for t,and solar masses for m.

We have shown that our model can reproduce the well established observed chemical abundance parameters of
the galactic disk in the solar neighborhood, the $^{9}$Be/H and $^{10+11}$B/H temporal evolution (see Casuso \& Beckman
(1997)), as well as that of D/H, $^{7}$Li/$^{6}$Li and $^{11}$B/$^{10}$B (see Casuso\& Beckman 1999).

It is well known that closed box models (with P(t)=0),of galactic
chemical evolution fail to reproduce several of the disk
constraints,most notably the metallicity distribution in the
disk,characterized by the low numbers of G dwarfs with low
metallicities.They also fail to reproduce the disk evolution of Be and
B vs. Fe, and models which give adequate fits to these observed data
sets require increasing infall to the disk of metal-free or metal-poor
gas (Casuso \& Beckman (1997)).We have therefore adopted the same
representation of the infall as was used in that paper:
\begin{equation}
P(t) = \frac {e{^{\lambda t}}}{M(t)}                              
\end{equation}

where M(t) is the total mass of the zone at time t,and $\lambda^{-1}$
is a time constant which must be in the range of a few Gyr.It is
notable that recent observations of abundances in the local
interstellar medium (Fitzpatrick (1996)) showing undepleted elements
with abundances significantly below solar, give support to the idea of
steady dilution by infall of non-enriched gas to the disk.  Including
the global effect of depletion as in the model of Casuso and Beckman
(1999) does not in practice yield significant improvements in the data
fit compared to non-depleted models, within the limits of error.

\section{Application of numerical modelling: incorporation of
$\alpha$+$\alpha$ by GCR in the ISM.}

A seminal early paper describing the physics of light element
production by spallation and fusion reactions,authored by
Meneguzzi,Audouze and Reeves (1971) has been the basis of much of the
intervening work in the field.These authors showed that nuclides of
light elements can be produced by spallation during collisions of
galactic cosmic ray (GCR) protons and alpha particles with nuclei of
C,N and O in the interstellar medium (ISM),and also by CNO in the GCR
colliding with protons and alphas in the ISM.The contribution of the
latter set of reactions to the production of Li,Be and B nuclides by
spallation has been estimated to be some 20$\%$ of the former
(Meneguzzi and Reeves 1975),and this
 should not have been very different in the past,assuming that the
 cosmic ray composition reflects that of the ISM.This has meant that as
 far as spallation is concerned we needed to calculate in detail only
 the former reactions,taking the latter into account by proportion.A
 further source of light element nuclides,whose importance has been
 recognized more recently,is the production of $^{6}$Li and $^{7}$Li by
 fusion reactions between GCR alpha particles and those of the ISM.The
 relative importance of this mechanism may have declined somewhat,as
 the abundaces of CNO in the ISM have grown relative to that of
$^{4}$He (although the abundance of the latter is still overwhelmingly
greater), but in the early phases of the disk it was certainly an
important mechanism (Montmerle,1977,Steigman $\&$ Walker,1992),and as
we will see,it must still play a major role today.

In this work we have used the standard expression for the production of
light element nuclides by GCR protons and alphas in the ISM:
\begin{equation}
\frac {dY{_k}}{dt} =
{\sum}Y{_j^{ISM}}(t){\sum\int}F{_i^{GCR}}(E,t){\sigma_{ij}^k}(E)dE     
\end{equation}
where Y$_j$(t) are the abundances,by number,of the various species,and
j refers to $^{12}$C,$^{13}$C,$^{14}$N,$^{16}$O or $^{4}$He,k refers to
$^{6,7}$Li and $^{9}$Be,$^{10}$B or $^{11}$B,and the variable i refers
to GCR protons or alphas.F$_i^{GCR}$(E,t) is the interstellar GCR flux
spectrum,and $\sigma_{ij}^k$(E) is the cross-section for each reaction
i+j$\longrightarrow$k,which has a corresponding energy threshold
E$_T$.The
quantities Y$_j^{ISM}$(t) are computed from the galactic chemical
evolution models described in section 3.They are thus observationally
constrained,and we have reasonably good estimates of their evolution
with t during the disk lifetime.The spallation and fusion cross
sections $\sigma_{ij}^k$(E) are also well-known (see Read $\&$ Viola
1984, Mercer,Austin and Glagola 1997) within narrow limits of error.These cross sections show rather
similar global behaviour,starting from thresholds E$_T$ close to 10-20
MeV/nucleon,peaking somewhere between 20 and 70 MeV/nucleon,and
declining rapidly to a plateau above $\sim$100MeV/nucleon.There
is,however,a key
 difference between the $\alpha$+$\alpha\longrightarrow$$^{6,7}$Li
 reaction cross-sections and the remaining cross-sections:while the
 peak value for the former is some 500 times that of the plateau
 value,for the latter the corresponding ratio is only 5 to 10.This
 implies that while for the processes that can give rise to
 $^{9}$Be,$^{10}$B and $^{11}$B the whole of the energy spectrum of the
 GCR,up to the GeV range,comes into play,for the $\alpha$+$\alpha$
 process only the lowest energy particles,those with less than 100
 MeV/nucleon,yield significant $^{6,7}$Li.(Of course a fraction of
 $^{6,7}$Li is indeed formed by spallation in the higher energy
 range,but none of the
$^{9}$Be or $^{10,11}$B can be formed via the $\alpha$+$\alpha$
process).This dichotomy has important consequences for the
observationally very different time
dependence,and hence the metallicity dependence,of $^{6,7}$Li on the
one hand,and $^{9}$Be/$^{10,11}$B on the other.

In order to try to reproduce the evolution of the light elements it is
clear that we need estimates of the current energy spectrum of the GCR
component,of its flux,and of how these parameters have varied with
time.For the present epoch the magnitude and spectral shape of the flux
of GCR particles reaching the earth are fairly well determined by
direct experiment for particles with energies higher than a few hundred
MeV/nucleon:those which are directly observed at the earth's orbit.
For these particles a spectrum of form
F(E,t$_0$)$\propto$E$^{-2.2}$ is found,up to a few GeV/nucleon and
F(E,t$_0$)$\propto$E$^{-2.6}$ at higher energies (Ip and Axford
1985).However at lower energies the GCR spectrum must be demodulated to
take into account the blocking effects of the heliosphere.This has been
a well-known cause of difficulties for light nuclide production
theory(Meneguzzi et al. 1971;Meneguzzi $\&$ Reeves 1979;Reeves $\&$
Meyer 1978),and the problem of determining the spectral dependence and
the amplitude of the unmodulated interstellar component of the GCR
spectrum below 100 MeV/nucleon still lacks an entirely acceptable
solution.In the present modelling exercise we follow Reeves $\&$ Meyer
(1978) in taking a most probable value for solar demodulation of 5,for
the whole GCR
spectrum,and supplement this with a further mean factor of 15 for the
particles with energies having energies E$\leq$100MeV/nucleon,which is
consistent with the estimates made by McDonald et al. (1990) and
McKibben (1991) from observations of the low energy component of
$\alpha$ particles out to a heliocentric distance of 43 a.u. with
Pioneer 10 and 11.More recent studies do not claim major improvements
here,because no direct measurement beyond the heliopause has yet been
made.To summarize,we take the spectral dependence of the flux to be
proportional to (E+E$_0$)$^{-\lambda}$,where E$_0$ is the rest energy
of the proton,and $\lambda$ takes values of 2.6 below
0.1GeV/nucleon,2.2 between 0.1 GeV/nucleon and 1GeV/nucleon,and again
2.6 at energies higher than this,and in
 these we follow previous studies on light nuclide production by Walker
 et al. (1985) and by Steigman $\&$ Walker (1992).Finally we normalized
 the flux to match the constraint imposed by the measured GCR proton
 flux for energies greater than 0.1 GeV/nucleon:12.5
 cm$^{-2}$s$^{-1}$GeV$^{-1}$nucleon$^{-1}$ (Gloeckler $\&$ Jokipii
 1967),together with a ratio of the $\alpha$/p fluxes of
 0.15,consistent with the observations of Gloeckler $\&$ Jokipii (1967)
 as re-examined by Webber $\&$ Lezniak (1974).Of course, one must
 invoke (as must all models invoking GCR reactions to produce the light
 elements) effective magnetic confinement of GCR's in the Galaxy in
 order to obtain the required high absolute fluxes of GCR protons and
 alphas.

In the present work we are concerned with Li production, but in the
models we have included depletion for the gas which has been processed
into stars (we have assumed here that stars which expel their gas into
the ISM have completely depleted their Li so that this expelled gas has
zero Li abundance), and the "{\it impoverishment}" due to the infall of
gas to the disk from the halo: we assume that this gas has the
initial,i.e. the primordial, Li abundance.However the inclusion, or
not, of these effects, influences the model results, i.e. the effective
production curve, significantly only in the range
-0.2$\leq$[Fe/H]$\leq$+0.2, where the effect of infall has been to
cause a slight fall in the observable Li abundance.

One of the arguments of the present paper rests on a correct
understanding of how the low energy GCR particle flux has developed
with time.Specialists in cosmic ray physics have previously proposed
that, since the measured abundances of GCR nuclides show a
dependence on the first ionization potentials of the parent atoms,the
principal sources of these particles must be the atmospheres of
relatively low mass,relatively cool stars (see Cass$\acute{e}$ $\&$
Goret (1978),Meyer (1985,1993)).The consequences of this for the time
dependence of the galactic GCR flux in the range of energies required
to produce the light element nuclides,and specifically for those below
100 MeV/nucleon
which participate in the $\alpha$+$\alpha$ fusion reaction,is an
increase of the flux at later times due to the accumulation of stars
with lifetimes comparable to that of the disk.In the seminal study of
Meneguzzi,Audouze $\&$ Reeves (1971) and also in the careful
re-examination of light element abundance production by Walker et al.
(1985),the zero-order assumption was made of a GCR flux constant with
time.A number of other workers in the field (Reeves $\&$ Meyer
1978,Mathews et al.1990) used a time-varying scheme in which the GCR
flux F(E,t) is proportional to the supernova rate,SNR(t),which in turn
was set proportional to the star formation rate SFR(t),in their
models.

In their study Prantzos,Cass$\acute{e}$ $\&$
Vangioni-Flam (1993) were well aware of the importance of the
time-dependence of the GCR flux,and also assumed that it followed the
supernova rate.However,in the present study we use the assumption that
this flux follows the expulsion rate of gas from stars,and we have in
fact varied the upper limit of the mass range from which expulsion is
considered.The delay entailed allows the model predictions to avoid one
of the main difficulties encountered by Prantzos et al.:that without
this delay there would have to have been an early sharp increment in the disk of Li
(in particular) against Fe,in the range of [Fe/H] between -2.0 and
-1.0,an increment
 which is not observed.This delay is due to the fact that although in
 the early disk there would have been a high SN rate,required to
 accelerate the GCR particles,the cumulative number of low and
 intermediate mass stars required to inject major quantities of He
nuclei (Meyer 1985)
 was still low.Both injection and secondary acceleration are required
 to yield MeV range GCR,and these conditions have been fulfilled
 simultaneously with increasing effect in the later disk,which explains
 the observed delay in the onset of disk Li production, even with
 respect to Fe (and {\it a fortiori} with respect to O).

In Fig 1 we contrast the evolution of the Li abundance in a typical
model in which the GCR flux is proportional to the SFR,with that in
models chosen from those we have applied in the present paper,in which
the flux is proportional to the gas expulsion rate from the stellar
population at a given epoch.The qualitative difference is evident,and
the relative reduction of the GCR flux in the early disk,compared with
more recent epochs,is clear.There are two further points about the
acceleration and propagation of the GCR which we should make here.As a
result of many studies over the past 30 years it is now a widely
accepted possibility that the majority of the observed particles in the
GCR flux have been accelerated in collisionless mode by shock waves
which originate in supernova explosions and propagate through the
dilute interstellar plasma
(Lagage $\&$ Cesarsky 1983,Blandford $\&$ Eichler 1987).In the model of GCR propagation by Prantzos et al. (1993) the high
energy part of the GCR flux spectrum is modulated according to the
escape length of the particles as a function of their energy;this
effect has changed with epoch,in such a way that the current spectrum
has a greater slope than the spectrum at early Galactic epochs.The low
energy fraction of the GCR flux has remained,however,virtually
unaffected by this change,suggesting that the evolution of the Li
production rate has been due rather to the time variation of
 alpha-particle density than to the variation of the spectral
 index.Secondly we must emphasize that provided there has been {\it
 sufficient} volume occupied by SN-affected ISM,the flux of low energy
 $\alpha$-particles will depend principally on the population density
 of the injectors:low-mass stars,rather than the SN remnants which
 accelerate them.A final point here is that low energy GCR's may in
 fact be accelerated by wind termination shocks due to stars of the
 full mass range.This process has been invoked by Rosner and Bodo
 (1996) to explain the diffuse non-thermal Galactic radio
 emission.Clearly acceleration via this process is not proportional to
 the SN rate but to the cumulative gas expelled by all stars present at
 a given epoch.Although taken alone it does not lead to sufficient
 delay to explain the abrupt rise in the Li-Fe envelope it is, a
promising mechanism for $\alpha$ acceleration in the context of the
Li abundance observations.

Here we should allude to the as yet not fully resolved question of the
origin of cosmic rays.In early models of SN shock theory, the thermal
gas in the ISM was regarded as the reservoir of seed particles which
can became cosmic-ray nuclei.But this clashes with the source
composition of the GCR (Meyer 1985).To solve this problem one needs to
invoke an injection of suprathermal ions.There are two main types of
scenario here: one assumes that the observed local flux of GCRs has
its origin in supernovae accelerating their own ejecta, and the other
assumes an origin in the atmospheres of intermediate and low mass
stars (for discussions see Meyer et al. 1997,Ellison et
al. 1997,Ramaty et al. 1997,Ramaty et al. 1998, Higdon et
al. 1999).While the early Galactic beryllium data suggest production
by cosmic rays originating from SN accelerating their own ejecta, the
observed composition of the cosmic-ray source material reflect a
correlation with first ionization potentials, leading to the
suggestion that cosmic-ray source material originates in the
atmospheres of stars.As evidence for this, we know that the abundances
of elements with low first-ionization potentials are enhanced in the
solar corona and in solar energetic particles, suggesting that similar
shock acceleration in low-mass, cool stars could provide a particle
injection source for acceleration by supernova shocks in the ISM.Both
origins (SN or low-mass stars) have many problems as complete
explanations of the origin of GCRs with energies greater than 1 GeV
per nucleon (see Ramaty et al. 1998). We cannot consider as
coincidental the similarity between the GCRS composition and that of
the solar corona which is biased according to first ionization
potential, and we must take very seriously the asertion of Ellison et
al. (1997) that "in the outer solar atmosphere the solar coronal gas,
the solar wind, and the $\sim$MeV solar energetic particles have
undoubtedly a composition biased according to FIP", together with the
fact that the hydrogen and precisely helium are not well fitted by the
alternative model of Meyer et al. (1997) and Ellison et al. (1997)
based on volatility and mass to charge to explain the GCRS.Also, we
must note that the cosmic-ray electrons have very different spectra
from that of the nuclear species at GeV energies, and may, in fact,
have entirely different origins (Berezinskii et al. 1990).In addition,
atomic collisions of low-energy ions (corresponding to a distinct
low-energy cosmic-ray component) produce characteristic nonthermal
X-ray emission.On this point Tatischeff et al. (1999) have shown that
a distinct Galaxy-wide low energy cosmic-ray component could account
for the hard component of the Galactic ridge X-ray emission in the
0.5-10 keV energy domain.Also, one must note the different behaviour
of helium and hydrogen data with respect to the other GCR nuclei when
the energy of these particles is increased , as inferred from the
observations in the solar corona, in the solar wind, in the solar
energetic particles and in the GCR (see Meyer 1985, Meyer 1993).One
can see how He and H abundances decrease systematically as the energy
increases, while the abundances of the other nuclei remain
invariant.All these considerations point to an origin for the
low-energy $\alpha$ particles which optimize Li production, which
could be quite different from that for the GCR nuclei at higher
energies.In our coherent scenario of chemical evolution for the
Galaxy, we point to the origin in low-mass stars of the low-energy
(those below 0.1 GeV per nucleon) $\alpha$-particles of GCRs,
consistent with the fact that the Li production cross section for the
$\alpha$-$\alpha$ fusion reaction falls very steeply outside the
energy range between 0.01GeV/nucleon and 0.1 GeV/nucleon (see
e.g. Ramaty et al. 1997).

In order to quantify our model, we will account for the energy needed
for the $\alpha$+$\alpha$ fusion reaction be coherent with the energy
supplied by the intermediate mass stars in a range when the Li
production is efficient.Because our concern is essentially with
alphas, and because the $\alpha$+$\alpha$ fusion reaction which yields
Li has a distribution of measured cross sections which is very low
outside the 10-100 MeV/nucleon range, we calculate the energy budget
by integrating the flux in that range.So, the energy per SNII will be
the total GCR energy in this range during e.g. 10$^{8}$yr., taking 0.1
cm$^{-3}$ as an average current He abundance in GCRs , and a section
corresponding to the 500pc radius taken here as the size of the
selected circumsolar volume, all divided by the number of SNII needed
by our numerical model ($\sim$13000 during each 10$^{8}$ yr.):

\begin{equation}
Energy/SNII=5{\cdot}15{\cdot}{\int_{0.01}^{0.1}}E(E+E{_{0}}){^{-2.6}}k{\cdot}dE{\cdot}10^{8}{\cdot}0.1{\cdot}{\pi}500{^{2}}/13000
\end{equation}
with E$_{0}$=0.938GeV, 5 is the solar demodulation factor and 15 a factor required to give a good fit to the data, and k is the normalization constant which is obtained fitting the observed GCR flux of alphas at energies above 0.1GeV:
\begin{equation}
0.15{\cdot}12.5cm{^{-2}}s{^{-1}}GeV{^{-1}}n{^{-1}}={\int_{0.1}^{1}}k{\cdot}(E+E{_{0}}){^{-2.2}}dE+{\int_{1}^{\inf}}k{\cdot}(E+E{_{0}}){^{-2.6}}dE
\end{equation}
From this one obtains the energy per SNII needed for our model to
produce the Li observed in the disk by alphas, of 2$\cdot$10$^{50}$
ergs. which is in very reasonable agreement for energy available from
SN model estimates.Taking the IMF used here, we also obtain that the
energy per star of intermediate mass (1-3 M$_{\odot}$) needed for the
alphas produced in this stars, is of 10$^{49}$ ergs, very reasonable
for the helium flash or coronal mass injections.

The net accelerating power of the OB star enviroment in the local
spiral arm has not,in this model, varied by more than a moderate
fraction during the disk lifetime.  The fact that the model produces
some twice as much local GCR flux at the present epoch as at the
begining of the disk is due entirely to the accumulation of particles
(H,He,C,N,O) emitted at low energies from low mass stars,and not to
any substantial change in the net efficiency of the SN mechanism which
subsequently accelerates them,and which has been present constantly
throughout the disk lifetime.

This scenario is consistent with other parameters such as the
lifetimes of $\alpha$-particles in the ISM.First we must note that due
to the very high temperatures needed to deplete $^{4}$He, efficient
destruction occurs in stellar interiors.However, the
$\alpha$-particles can also disappear in principle due to fusion
reactions with other $\alpha$s,$^{12}$C,$^{14}$N and $^{16}$O in the
ISM, leading to
$^{6}$He,$^{6}$Li,$^{7}$Li,$^{7}$Be,$^{9}$Be,$^{10}$Be,$^{10}$B,$^{11}$B,$^{10}$C
and $^{11}$C.These latter reactions have cross sections below 100mb,
and so, taking densities of the ISM below 10$^{5}$cm$^{-3}$, the
lifetimes are greater than 10$^{12}$ yr, i.e. greater than the age of
the Universe.

On the other hand, the flux of $\alpha$-particles in the range of
energies concerned here, i.e. between 10MeV/n and 100 MeV/n, could
comes from the $\alpha$-particles produced in later type stars at
energies below 2.5MeV/n.  In the mass range
0.5M$_{\odot}$$\leq$M$\leq$3M$_{\odot}$ the central regions become
degenerate and the triple-alpha reaction ignites via the violent {\it
helium flash}.The central energy-generation rate at the peak of this
helium flash exceeds 10$^{13}$ times that in the center of the Sun
causing the well attested expansion to the giant phase before the bulk
of the He has been consumed.For these stars the expulsion of He nuclei
into the ISM occurs with much greater efficiency than for the less
accelerated transformation to the giant phase accompanied by the
burning of the He which occurs in stars with higher masses.Another
realistic possibility to explain the required low energy alpha flux is
that the ions can be injected (at MeV energies) via coronal mass
injections (mainly from the coronae of dMe and dKe dwarfs, by far the
most numerous stars in the Galaxy) (Shapiro 1999).

One way of seeing the problem is via a two-stage scenario similar to
that of Meyer (1985), which assumes the OB associations as the best
sites of production of $\alpha$-particles of GCRs; there, a large
number of later type stars are being formed together with a few
short-lived massive stars; the former have a very high surface
activity owing to their youth and should emit lots of suprathermal
particles, while the latter provide stellar wind and SN shock waves
within their few 10$^{6}$ yr lifetime; so injectors and high energy
accelerators are closely linked in space and time.Energies as low as
0.01-0.1MeV/n are sufficient for suprathermal particles to be
accelerated much more efficiently than the thermal gas.Particles with
energies below 2.5MeV/n undergo significant coulomb energy losses,
which brake and thermalize the particles, impeding $\alpha$+$\alpha$
fusion production of Li.However, the time for thermalization is
inversely proportional to the density of the medium in which
suprathermal particles propagate.The $\alpha$-particles propagating in
dense clouds, thermalize within 10$^{4}$ yr, but in the diffuse hot
interstellar medium (HIM) where n$_{H}$$\sim$3.10$^{-3}$cm$^{-3}$, the
thermalization time is several times 10$^{6}$ yr.We can then assume
that the mechanisms already proposed by Meyer (1985) can
operate.Ionized media confine energetic particles, so that
suprathermal particles emitted in the HIM will not in general traverse
any neutral, dense medium.In dense cloud complexes later type stars
can form continuously while OB star formation will disperse the
complex rapidly; and Meyer (1985) brings out the possibility of
reacceleration of suprathermal particles emitted by young late type
stars having migrated into the HIM just nearby the cloud complex.

But one might not in fact need a two-stage scenario if the
intermediate stars produce and also accelerate alphas to energies in
the range 10-100MeV/n, where the coulomb energy losses are not so
efficient.

Another possibility is that the $\alpha$-particles needed come from
the so called "anomalous" component of He nuclei which is observed
precisely at the range of low energies here considered (below 100
MeV/n).These He nuclei had reported as an unusually flat helium
spectrum, apparently unrelated with GCR spectrum (Webber 1989).In
fact, the solar modulation effects on this helium anomalous component,
as observed on the Pioneer 10 spacecraft at $\sim$40 AU from the sun,
between 1985 and 1987 (when the solar modulation reached its minimum)
show a change of a factor $\sim$100 in the intensity of these
particles between 10 and 20 MeV/n as they rapidly emerge from the
background of low energy GCRs (Webber 1989), in good agreement with
this work where we need a total modulation of 5$\cdot$15=75 over the
"normal" GCRs.

We can now summarize the reasons why the present family of models gives
an adequate prediction of the observed Li vs. Fe evolution curve.In
those older models where a constant GCR flux was used (e.g.
Walker et al. 1985) which is in fact not too bad a first-order
approximation to the time-delayed evolution for the low energy flux
which we obtain here,the importance of the $\alpha$+$\alpha$ fusion
reaction was not realized.In more recent models,on the other hand,where
$\alpha$+$\alpha$ reactions have been well included (Steigman
$\&$ Walker (1992),Steigman (1993),Prantzos et al. (1993)), sophisticated time-evolution schemes for the GCR have been
used, which unfortunately do not explore the delayed contribution
of lower mass stars to the low energy GCR spectrum.In the former models
normalized to give the correct Li abundance at, say,
[Fe/H]$\simeq$-1.5, there was simply not enough contemporary Li
produced and in the latter,while it would be feasible to attain the
contemporary abundance value: log N (Li) $\sim$ 3,this would entail
abundances of Li at [Fe/H] $\sim$ -1 which are too high by more than
half an order of magnitude.In the next section we show that our models
yield results in much better agreement with the observations.

\section{Predictions of the family of numerical models: comparison with
observations.}

In the present section we present the results of our modelling
exercises.We have already shown in Fig. 1 the observations that we have
set out to model.We began with data which we ourselves reported
(Rebolo,Molaro $\&$ Beckman 1988) because of the ready availability of
the complete data set, to which we have added a newer and extensive set
of results from Spite (1996).In Fig 1 we show observations of Li vs. Fe
abundances over a wide range of surface temperatures and
metallicities.The assumption we will make in interpreting these data is
that the upper envelope shows,essentially,the evolution of Li with
Fe,while the points which fall below the envelope refer to Li depleted
in the individual objects observed.

In Fig 1 we see that the Li abundance remained essentially constant
during the halo period (in which the Fe abundance was evolving from its
lowest values to around -1.5) and then began to rise.The approximate
plateau at low [Fe/H],the "{\it Spite}" plateau,corresponds roughly to
primordial Li,while the later rise represents the presence of galactic Li
production modulated by any averaged depletion which may take place.A
key  observational result is the rather abrupt rise of the Li envelope
at [Fe/H]$\sim$-0.2, to an essentially constant value between -0.2 and
+0.2.This second "{\it plateau}", although it is not as clear as the
"{\it Spite plateau}" appears to be consistent with our general model
predictions as one can see in Fig. 1.This plateau is a natural
consequence of the "{\it loop back}" in abundance of Li already shown
in Casuso \& Beckman (1997) to occur for the Be and B disk abundances
(where it appears more distinctly because stellar depletion is much less
important than for Li), and is due to the increase with time in the
infall of gas to the disk, which dilutes the Li abundance and more so
the Fe abundance, reducing the latter in recent epochs from a broad
peak attained several Gyr ago.

The model shown in Fig 1,in which the GCR flux was held proportional to
the gas expulsion rate from the whole stellar population represents a
first approach using delayed $\alpha$+$\alpha$ as the principal source
of Li,and its relative success is encouraging,but in using gas expelled
from the whole stellar mass range it does not try to take into account
the fact that the low energy component of GCR,responsible for the
$\alpha$+$\alpha$ process which we are postulating as the principal source of Li, may well originate mainly in lower mass
stars.The most direct way to do this is to place  upper stellar mass
limits on the expulsion rate of gas for which the GCR flux,at
 each epoch,is deemed proportional.Curves (ii) and (iii) show the
 results of allowing the GCR flux which enters the $\alpha$+$\alpha$
 reaction to be proportional to the gas expelled per unit time from
 stars of all masses and up to mass limit of 3M$_\odot$ respectively.It
 is clear that the model with the upper limit of 3M$_\odot$ gives a
 much better fit to the observed envelope shown here, and goes far in
 demonstrating the need to include the implied time delay in the
 build-up of the GCR flux in the relevant energy range.The curve for
 the 2M$_\odot$ upper limit shows a good fit to the form of the
 observations but yields rather low Li production.

Even given the observational uncertainties in the Li-Fe dependence
we can use these data and these models to constrain broadly the
stellar mass range which serves as a significant source of low energy
GCR $\alpha$ particles.The fact that models with an upper mass
restriction give better fits to the data is itself an argument in
favour of intermediate and lower mass stars as the principal sources of
the GCR flux at low energies.

One
further
note should be added.Our evolutionary models were designed to account
for the G-dwarf metallicity distribution, and B and Be evolution in the
disk.In fact the latter data begin to be very weak statistically for
[Fe/H] greater than +0.1, which is where the previous model predictions
terminate.In Fig. 6 we show the result of a modified model where the
upper limiting Fe abundance is +0.3.This is an $\alpha$+$\alpha$ model
with the relevant GCR flux proportional to the gas expelled from stars
with
masses $\leq$3M$_{\odot}$, and it also accounts well for the
observations.We must note, however, that the systematic uncertainties
admitted by the observers for the stars with high Li abundances and high
Fe abundance, are in the sense of requiring lower [Fe/H] (0.1 instead
of 0.3 or 0.4) (see e.g. Boesgaard \& Tripicco (1986)).

\section{Discussion: alternative sources of disk lithium.}

We have used evolutionary models for the galactic disk in the solar
neighborhood to re-examine the theme of Li production.In the first part
of this paper we presented analytical and semi-analytical models, with
their simplifying assumptions and limitations, and showed that models incorporating a delay in Li production compared with that of Fe give a fair
description of the observed evolutionary history of Li,using the Fe abundance as
a reference parameter.In the subsequent sections we showed that
numerical models with infall of non-enriched gas during the disk
lifetime give good fits to the observations, the best fits coming from
models where the infall has shown a tendency to increase (see Fig.5).One
consequence of this has been the non-monotonic evolution of the
metallicity with time:[Fe/H] has grown to somewhat greater than solar
values,and then fallen back slowly.This circumstance yields metal-metal
plots with a characteristic fold-back close to solar metallicities.The
Li production depends in fine detail,but not in principle or in broad
trend on this form of the infall model.

Applying this general evolutionary scheme to Li production we have
introduced one novel assumption which proves capable of resolving
the hitherto difficult to reproduce Li-Fe curve.This assumption is that
the part of the GCR flux responsible for the production of $^{7}$Li and
$^{6}$Li by the $\alpha$+$\alpha$ fusion process:the low energy flux
(at least) is emitted principally by intermediate mass stars,an
assumption well supported in the literature on GCR production (see e.g.
Meyer 1985,Meyer 1993 and references
therein).The novelty which this introduces into the models is that the
production rate of $^{7}$Li and $^{6}$Li is then constrained to follow
not the SFR,but the rate of net expulsion
of gas from stars within a mass range whose upper limit becomes an
independent variable in the modeling scheme.This assumption leads to
the delayed production of Li in the models,in good agreement with the
observations.

The concept of late-time production of Li is not,of course,newly
introduced within the present model,even though its use with a GCR
source here is original.Other mechanisms for producing Li associated
predominantly with lower mass objects might in principle be able to
satisfy the observational constraints of the Li-Fe envelope of Fig.
1,and a number of these have been suggested.Sites which have been put
forward as serious candidates for a major fraction of galactic Li
production include novae,neutron stars,stellar mass black holes, red
giants, AGB stars, and carbon stars,all of which might satisfy the
constraint of delaying the Li production with respect to that of Fe
and also processes in supernovae,which satisfy this condition with
greater or lesser degree of difficulty.It is not possible to dismiss
these sources but the fact that they each fare some
difficulty:theoretical or observational,tends to reinforce our view
that GCR rather than stellar production of Li has in fact
predominated.

Results of earlier work by Arnould $\&$ Norgaard (1975) on
novae,followed by the more detailed study of Starfield et al. (1978)
have been more recently called into question by Boffin,Paulus $\&$
Arnould (1993) using new reaction cross-sections.The latter authors
conclude that it is much more difficult to produce a significant
quantity of Li in novae than previously predicted.A search for lithium
in late-type companions of several dwarf and classical novae has not
yielded detections (Martin et al. (1995))

Measurements of Li in the youngest stars,in some of whose atmospheres
production has been postulated to occur,tend to yield abundances close
to the "{\it canonical}" value for moderately young stars,of log N (Li)
$\simeq$ 3.2 or 3.3,and do not seem to show sufficient Li to be strong
candidates for major galactic production.Martin et al. (1992) showed
that in some cool companions of hot (i.e. young) stars,where simple
atmospheric model analysis could appear to show Li abundances of up to
3.7,a more careful NLTE study yields upper values of 3.4,with most
objects falling below this.Similar results have been obtained by
Duncan (1991),by Magazzu et al. (1992) and by Martin et al. (1992) for
T Tauri stars.Here again careful analysis reduced apparently very high
values of the Li abundance in some objects to values well within the
range of the normal young stellar population.

There do appear to be giants especially over-abundant in
Li,notably a sub-set of the C-stars (Abia et al. (1991)),and a fraction
of normal K-giants (De la Reza \& Da Silva (1992)).Observations here
are still few,and conclusions made somewhat more difficult by the
convective tendency of giants to deplete Li (this of course strengthens
any argument in favor of such stars being Li sources if strong Li
absorption lines are seen in their spectra).In the most extensive sets
of observations of field giants,however,very few indeed have
particularly strong Li abundances (Brown et al. 1989).Thus if certain
types of giants are important Li producers,since they are few,and
therefore need to be strong sources,while sparsely distributed,one
might expect more scatter in population I Li than is
observed.Nevertheless production in giants remain a possible source,and
the main argument we offer against its being the main source can only
be that the GCR model presented is able to account for the relevant
observations without a major extra Li contribution.

Similar consideration may be given to the postulated importance of Li
production in late-type companions to neutron stars and black-hole
candidates.In a paper on these objects by Martin et al. (1994),Li
abundances ranging up to 3.3 for Cen X-4 are detected
which might be increased if there is substantial overionization of LiI
due to UV and X-ray flux coming from the compact object.These authors
claim that since Li undergoes depletion by convection in late-type
stars,the presence of relatively high Li abundances in these objects
marks them as Li producers,and therefore candidates for major
enrichment of the galactic
disk.Here again while we see no immediate argument which can rule out
this possibility one may doubt that there are sufficient such
sources.Martin et al. (1994) put forward the idea that there may well
have been more
X-ray binaries,especially high-mass binaries,in the past but the curve
of Li as a function of Fe implies that the production mechanism should
not be associated with high mass objects.Nevertheless we are not in a
position here to claim that processes in X-ray binaries cannot be
responsible for a significant part of galactic Li production,only that
these appears to be no requirement for this as a major source.

In all the analytical approximations based on proportionality of Li
production rate to the gas expulsion rate by stars of low or
intermediate masses (equations (12) (14)) one can see
considerable resemblance to the observed lithium growth profile in the
zone of interest, i.e., with metallicities [Fe/H] between -1.0 and 0.0
(see Fig. 1 and Fig. 3).

We have used the $\alpha$+$\alpha$ process in the ISM as illustrative
of processes which follow the behaviour of stars of intermediate and
low masses, those whose lifetimes are long, and whose numbers in the
disk have therefore grown cumulatively with time.Any process with
equivalent time-dependent characteristics could, in general terms,
satisfy this global observational constraint.Production of Li in flares
of red giants (Cameron $\&$ Fowler 1971), or in novae (Arnould $\&$
Norgaard 1979,Starrfield et al. 1978), could also
satisfy, in principle, the criterion of delayed production because
there the rate of production would be proportional to the number of
stars with intermediate masses at each time (cumulative because of
their long
lifetimes) and not to the gas expulsion rate.But, as one can see in Fig. 4, where we plot the results
from numerical model in the case of GCR flux proportional to the
cumulative number of stars with masses less than or equal to
1M$_{\odot}$ (increasing this limit yields a decreasing rate of
accumulation of Li producers), the prediction, although not too bad, is
by no means as good as that obtained using gas expulsion by stars of
masses less than or equal to
3M$_{\odot}$, and is also in fact worse than that obtained using gas
expulsion of stars with masses less than or equal to 2M$_{\odot}$ (see
Fig. 4).

As far as mechanisms which depend on the presence of SNI are concerned,
two arguments appear to weaken their claims.One is that the Li-Fe
envelope in Fig. 1 is not linear, which would be the dependence if
the Li were either produced directly by the impact of SNIe on their
immediate surroundings, or by processes involving GCR in a wider volume
of space, produced by SNIe.The other is that the locally measured GCR
abundances favour an origin, for the lower energy particles at least,
in the thermal equilibria pertaining in the atmospheres of stars of
moderate mass, rather than in the extreme conditions of a supernova.However here again we argue in terms of probabilities rather than claming that this mechanism is excluded.

Another possibility is the production of Li in compact objects such as
neutron stars and black holes, but the time evolution of compact
objects would be proportional to the gas expulsion rate from all stars
(which is mainly that of SNI+SNII).This production is shown in Fig. 1
from numerical modelling, and one can see how this implies
overabundance of Li with respect to the observational constraints; in
fact its tendency is not too different from that of those GCR models
where the flux is proportional to the SFR.

Other possibilities for production of Li, such as in AGB stars or
carbon stars (Matteucci et al. (1995), Romano (1999)) are shown not to reproduce the
observations really as well as the delayed models, as one can see, for example, in Fig. 4.

\section{Conclusions.}

We have surveyed mechanisms of Li production in the disk, and
confronted them with the upper envelope of the Li-Fe observations,
which we have taken to represent the Li-Fe evolution curves in the
absence of stellar depletion for the individual objects observed.As a
result we can conclude that:

1)Mechanisms relying on SNIIe to produce Li cannot be excluded 
in at least an approximate explanation of the observations.This is true for production within the SNe
themselves, but also holds for GCR fluxes originating in SNIIe.

2)Mechanisms whose time dependence is that of the SFR give
either too much Li in the early disk or too little in the later disk.

3)Mechanisms which rely on SNIe to produce the Li (again either in
the immediate surroundings of the SNa or via a more generally
dissipated GCR flux originating in SNIe) predict that the disk Li
should grow proportionally to Fe, which does not appear to fit the observations.

4)An attempt to reproduce the results on the assumption that the
contemporary maximum abundance, logN(Li)$\simeq$3.4 is the true
primordial abundance, and that the Li-Fe envelope is a pure depletion
curve also fails by a wide margin.

5)Mechanisms which produce an increase in disk Li significantly
delayed with respect to that of Fe can explain the observations very well.

We have in this article explored one such mechanism: the production of
Li via $\alpha$+$\alpha$ fusion reaction in the ISM due to low energy
cosmic rays whose source of origin is the atmospheres of low and
intermediate mass stars.This mechanism has the virtue that these stars
have lifetimes comparable with that of the disk, so that their
collective gas expulsion rate has accumulated progressively throughout
the disk lifetime, leading automatically to a delay with respect to Fe
in the Li production curve.We have explained that even if the
acceleration of the GCR is due to SNe envelopes, the product of
injection rates and acceleration rates retains the delay implied by
the observations (further work on acceleration mechanisms such as that
due to stellar wind termination shocks is, however, well worth
exploring in this context).Support for the possibility of this
mechanism is provided by the observed similarity between the GCRS
composition and that of the solar corona which is biased according to
the first ionization potential, and we note in this context the
statement of Ellison et al. (1997) that "in the outer solar atmosphere
the solar coronal gas, the solar wind, and the $\sim$MeV solar
energetic particles have undoubtedly a composition biased according to
FIP", together with the fact that the hydrogen and precisely helium
are not well fitted by the alternative model of Meyer et al. (1997)
and Ellison et al. (1997) based on volatility and mass to charge to
explain the GCRS.These considerations permit an origin in an
environment close to thermal equilibrium, i.e. typical of stars of
moderate mass.We have incorporated the mechanism in an evolutionary
model of the disk previously demonstrated to be capable of accounting
well for the Be and B vs. Fe observations (Casuso \& Beckman (1997)),
and which gives a particularly good account of the G-dwarf metallicity
distribution in the solar neighborhood.The resulting Li-Fe plots
include very fair fits to the observed Li-Fe envelope.

We have included in this scenario a natural mechanism of differential depletion (Casuso \& Beckman 1999) operating within red supergiant envelopes, which can account for the observed D/H v. time and isotopic ratios of $^{7}$Li/$^{6}$Li and $^{11}$B/$^{10}$B v. time.

However we would not at this stage wish to rule out the possibility of other mechanism or mechanisms for disk lithium
production.The observational weight of the stellar Li abundances,
as we
have shown, does place some strong constraints on Li-production models.One of the clearest
conclusions we can draw is that the "high" value log N(Li)$\simeq$3.4
for the primordial Li abundance can be quantitatively rejected using
the Li-Fe observational constraint.The assignation of a value close to the "Spite
plateau" (Spite \& Spite (1982)) value: log N(Li)$\simeq$2.2 as
primordial is thereby strengthened.In this context the comprehensive
study by Thorburn (1994) of Li in halo stars, in which a contribution
to the plateau produced by the $\alpha$+$\alpha$ reaction due to
the halo GCR flux is shown to account well for the observed scatter and
slight rise in the Li abundance below [Fe/H]=-1.5, makes a suggestive
link with the disk model tested in the present paper.The importance of
the $\alpha$+$\alpha$ process has almost certainly been previously
underestimated in the disk, and the powerful constraint on evolutionary
processes and models implied by the Li vs. Fe observations has not been
adequately taken into account; it is these aspects of the lithium
puzzle which the present paper has been designed to expose.

{\large Note added in Proof:} Newly observations of Li and
$^{7}$Li/$^{6}$Li in ISM (toward o Per and $\zeta$ Per) by Knauth,
Federman, Lambert, Crane (Nature in press), give a variation in
$^{7}$Li/$^{6}$Li ratio (from near 2 which is the expected for Li
production from spallation or alpha-alpha fusion reactions purely, to
near 11 which is very similar than that of solar value), together with
very similar reported values for Li/H abundance (near 11x10$^{-10}$)
for the two clouds in contrast with the solar value of
20x10$^{-10}$).Also, the two clouds are near the star forming region
IC 348.  All of these data agree very well with our picture of
production of light elements in the ISM via GCRs (Be,B) (Casuso and
Beckman 1997) and via alphas of low-energy (Li).We explained this
variation (in fact a fall off) via a model in which the envelopes of
red-supergiant stars (so, star forming region) deplete differentially
$^{6}$Li and $^{7}$Li , and the increasing infall of non-depleted gas
with time (Casuso and Beckman 1999).And also, we explained in the
present article the decay on Li/H abundance from solar to actual ISM
due precisely to the depletion in star forming regions in addition
with the infall of non-enriched gas (see Fig. 4).So, we can explain
these data without the problem inh erent to the explanation by Knauth
et al., which point to the differential production of Li in the o Per
direction and in the $\zeta$ Per direction because of the higher flux
of cosmic rays in the o Per direction, while observations point to
almost the same total Li/H abundance.

\acknowledgments{\large Acknowledgments:}
We are happy to thank F. Spite for supplying his lithium abundance
data compilation, and for helpful suggestions, and E.L. Martin for
useful discussions.The anonymous referee made a number of valuable
suggestions which led to significant improvements in the paper.This
research was supported in part by grant PB97-0219 of the Spanish
DGICYT.

\newpage

\parindent=0pt

{\bf Figure Captions}

\vspace {0.5cm}

{\bf Fig.~1} a) Abundance of Li v. iron metallicity [Fe/H] in the solar
 neighborhood.Compilation from the work of the group of the present
 authors (Rebolo,Beckman,Molaro,1987,Rebolo,Molaro and Beckman,1988), and from Spite (1996).A
 conservative error bar is shown in the left upper corner.The upper
 envelope characterizes the stars least depleted in Li at the epochs
 implied by each value of metallicity.Stars below the upper envelope
 have suffered notable internal Li depletion.  In the present article
we model only the envelope for the disk,i.e. for [Fe/H]
$\geq$ -1.5.  b) Disc production curves of Li for three models in which
the low energy GCR flux responsible for the Li,via $\alpha$+$\alpha$,is
proportional to:(i)The star formation rate (dotted line),(ii)The gas
expulsion rate for the population of stars with all masses (dashed
line),(iii)The gas expulsion rate for the population of stars with
masses $\leq$ 3M$_\odot$ (full line).Each curve is normalized to give
the observed Li abundance of 2.4 at [Fe/H]=-1.3.The curves are shown in
comparison with the data.The axes in this figure, and in Figs. 3, 4 and 6,
are: for [Fe/H] the logarithmic abundance with the solar value
[Fe/H]$_{\odot}$=0, for Li the logarithmic abundance where logN(H)=12.
\\ \\

{\bf Fig.~2} a) Iron metallicity in the solar neighborhood as a
function of age.The curve show the prediction of our model.Observation
comparison points with error bars are from: Meusinger et al. (1991)
(crosses) and Twarog (1986) (crossed circles).b) The ratio oxygen/iron
as a function of the iron abundance.[O/Fe] v. [Fe/H] data are from:
Rebolo et al. (1994),Nissen et al. (1994),Israelian et al. (1998).The
full line corresponds to our chemical evolution model using the yields
of Fe theoretically calculated for stars with very low metallicities,
and the dotted line is for yields from stars with intermediate
metallicities.However, in the present paper we are critically
concerned with iron metallicities greater than -1.5 where there are no
major problems with the dispersion of data.\\ \\

{\bf Fig.~3} Normalized plots based on models embodying analytical
schematic approximations to Li production rates and their evolution
with Fe, to illustrate how some selected scenarios for the evolution
of disk Li are much less in agreement with the form of the
observations (the upper envelope of the points in Fig. 1a), than are
others.Points are the data as in Fig. 1.All curves are from analytic
approximations by mathematical functions with two free parameters
which are constrained to match the data envelope at [Fe/H]$\simeq$-1.0
(where the full disk initiates) and at [Fe/H]$\simeq$0.0.Long dashed
line: from eq. (17).Full line: from eqs. (6) and (19) which give a
very similar result.Dashed line: from eq. (14).Dotted line: from
eq. (12). \\ \\

{\bf Fig.~4} a) through d).Observations of Li vs. Fe compared with the predictions of
models taken from the literature, with parameters described in the text,
and also with models developed in the present paper.Graphs common to all
panels: Model due to Prantzos et al. (1993): dashed plus dotted
line.Model due to Matteucci et al. (1995): dotted line.Halo model due
to Casuso \& Beckman (1997): solid line for [Fe/H]$\leq$-1.3
only.Differential curves: a) $\alpha$+$\alpha$ model with GCR flux
$\propto$ gas expulsion rate by stars of masses $\leq$3M$_{\odot}$
(full line); model with GCR flux $\propto$ gas expulsion rate by stars
of masses $\leq$2M$_{\odot}$ (dashed line).
b) $\alpha$+$\alpha$ model with GCR flux $\propto$ gas expulsion rate
by stars with masses $\leq$3M$_{\odot}$, with exponentially increasing
infall (full line), with no infall (dashed line).  c) $\alpha$+$\alpha$
model with GCR flux $\propto$ gas expulsion rate by stars with masses
$\leq$3M$_{\odot}$ (full line); with GCR flux $\propto$ cumulative {\it
number} of stars with masses $\leq$1M$_{\odot}$ (long dashed + dotted
line).
d) $\alpha$+$\alpha$ model with GCR flux $\propto$ gas expulsion rate
by stars with masses $\leq$3M$_{\odot}$ (full line); model with no disk
production but with linearly time-dependent stellar depletion from a
"primordial" value of log N(Li)$\simeq$3.5 (long and short dashed
line). \\ \\

{\bf Fig.~5} Example showing the essential difference in GCR flux as a
function of
 time for two key models: proportional to SFR(t) (full line) and proportional to the gas expulsion rate by stars
 with masses less than or equal to 3M$_{\odot}$ (dotted line) against
 time, from our numerical model with increasing infall of gas to the solar
 neighborhood.The steps in the curves are unsmoothed constructs of the
 model due
 to the finite time intervals employed. \\ \\

{\bf Fig.~6} Extrapolated model curve which can account for Li
abundances observed with [Fe/H] greater than 0.1. \\ \\

\begin{figure}
\label (a)
\plottwo{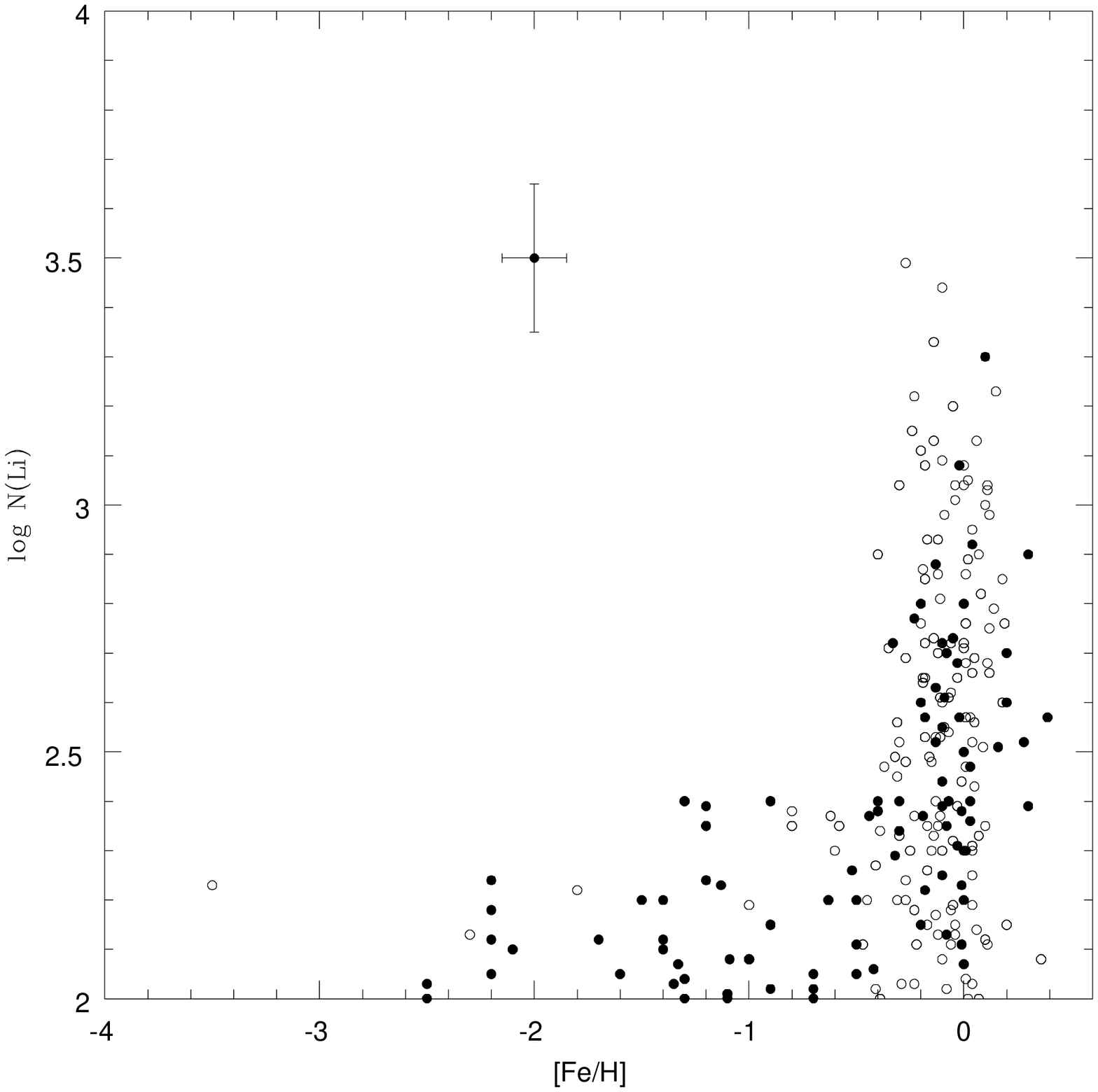}{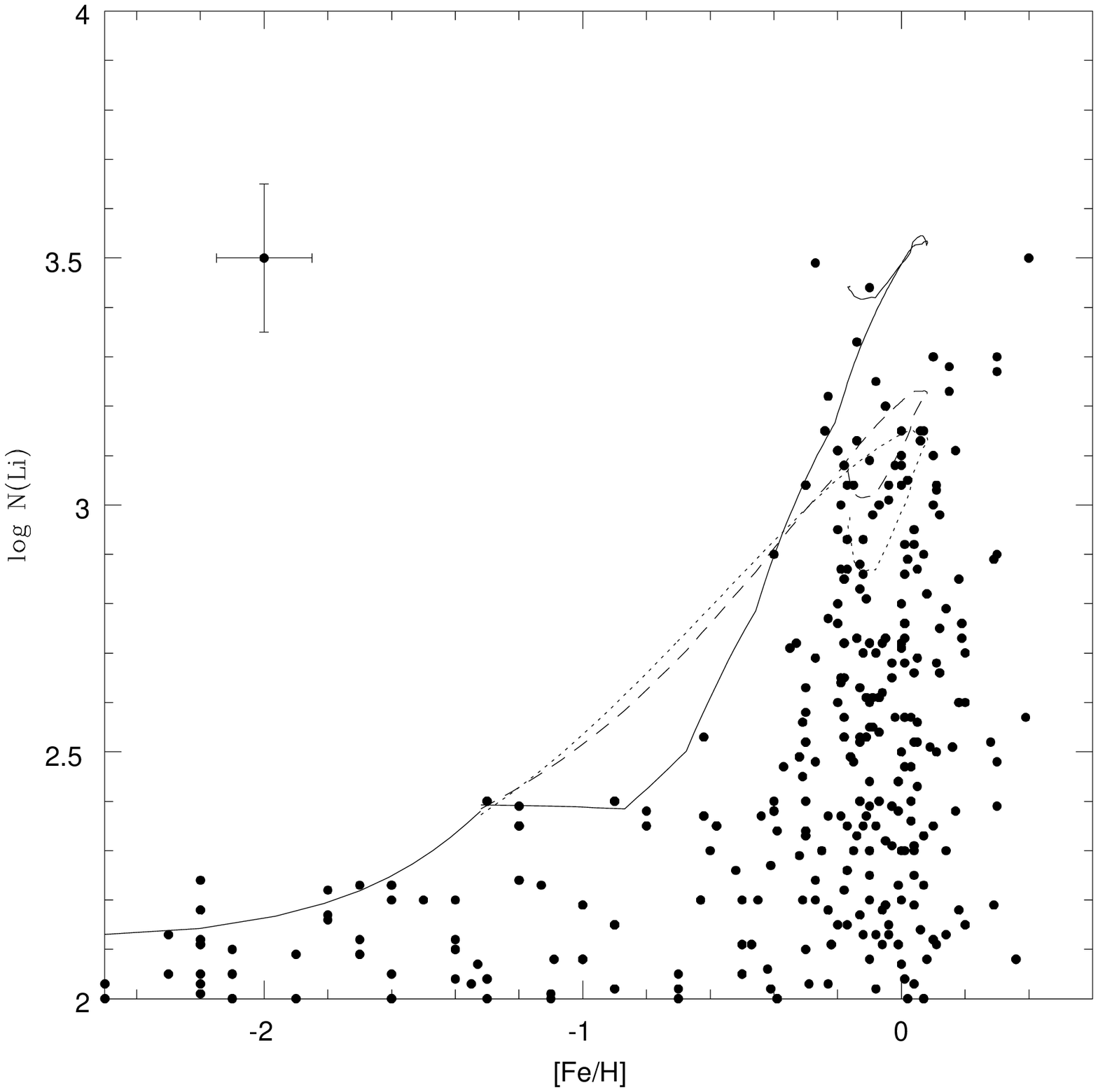}
\label (b)
\end{figure}
\begin{figure}
\label (a)
\plottwo{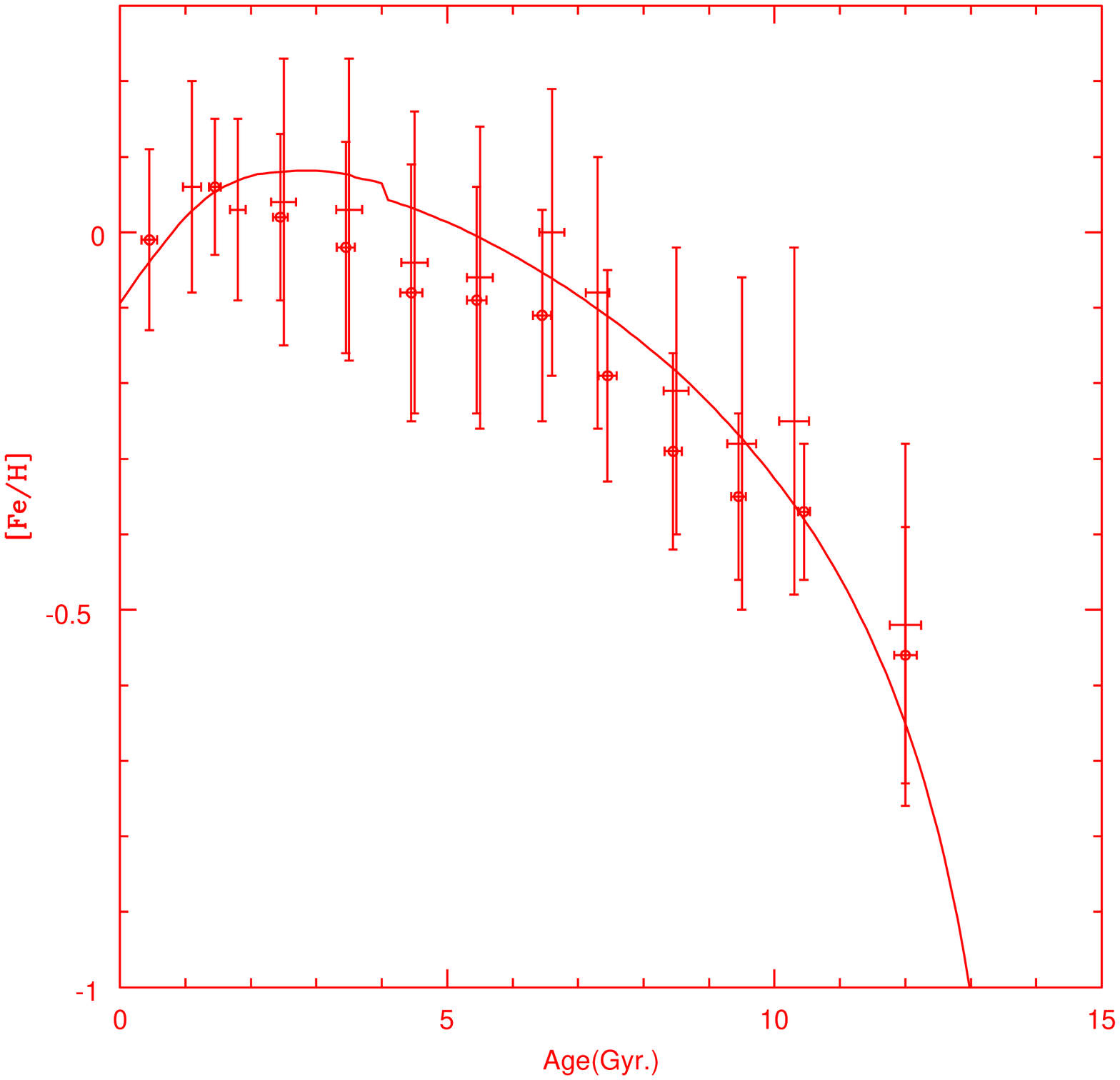}{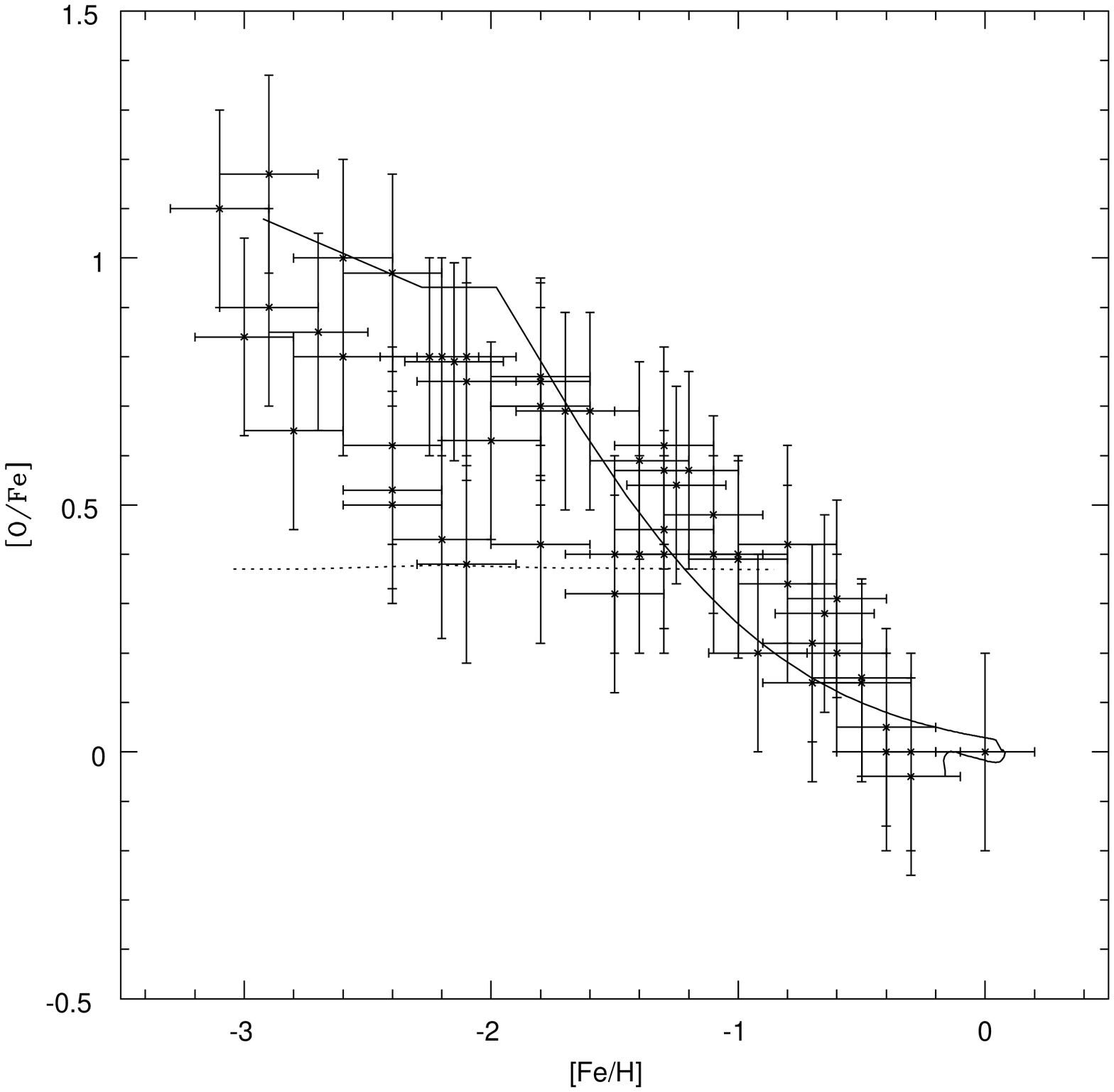}
\label (b)
\end{figure}
\begin{figure}
\plotone{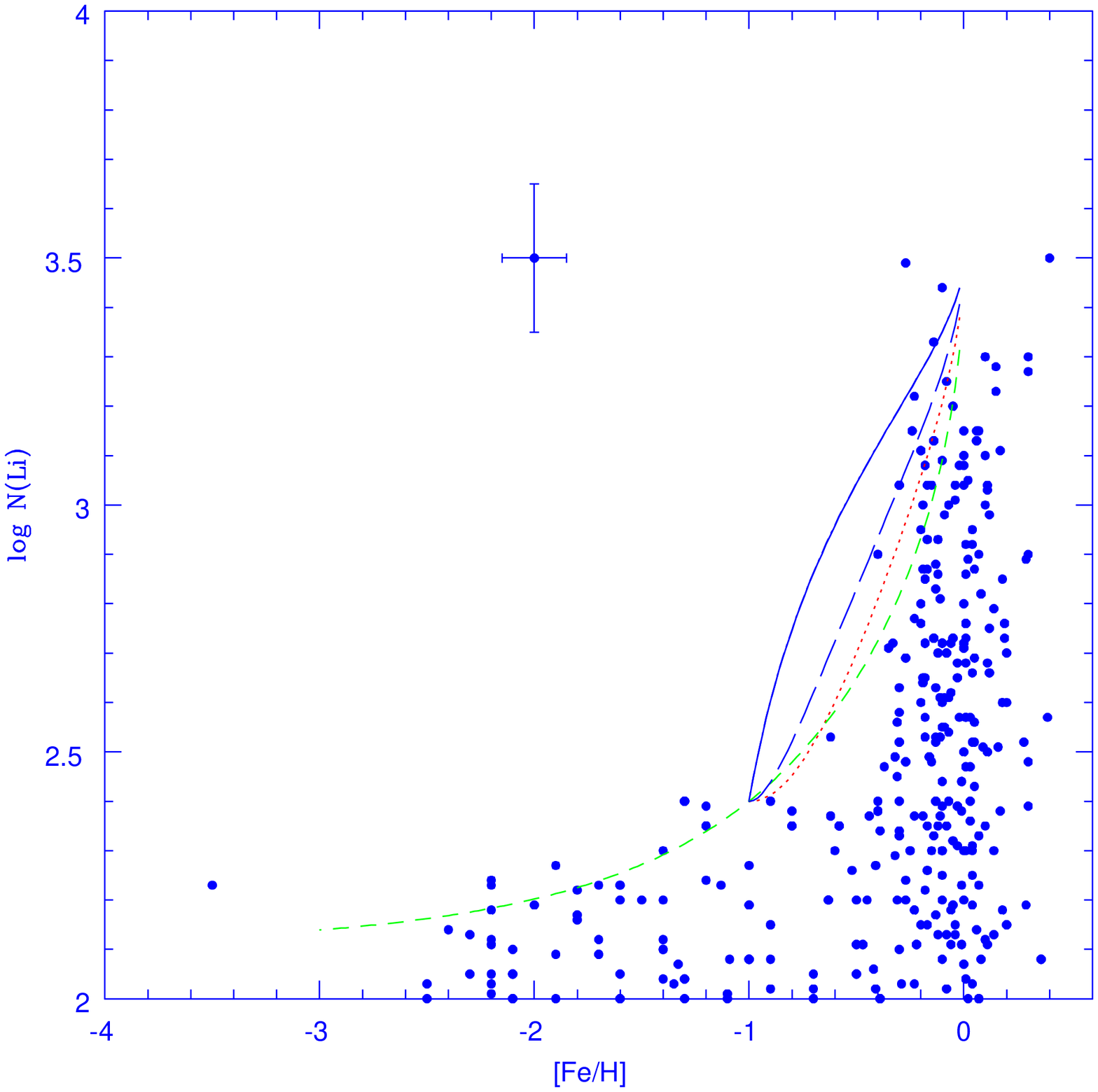}
\end{figure}
\begin{figure}
\label (a)
\plottwo{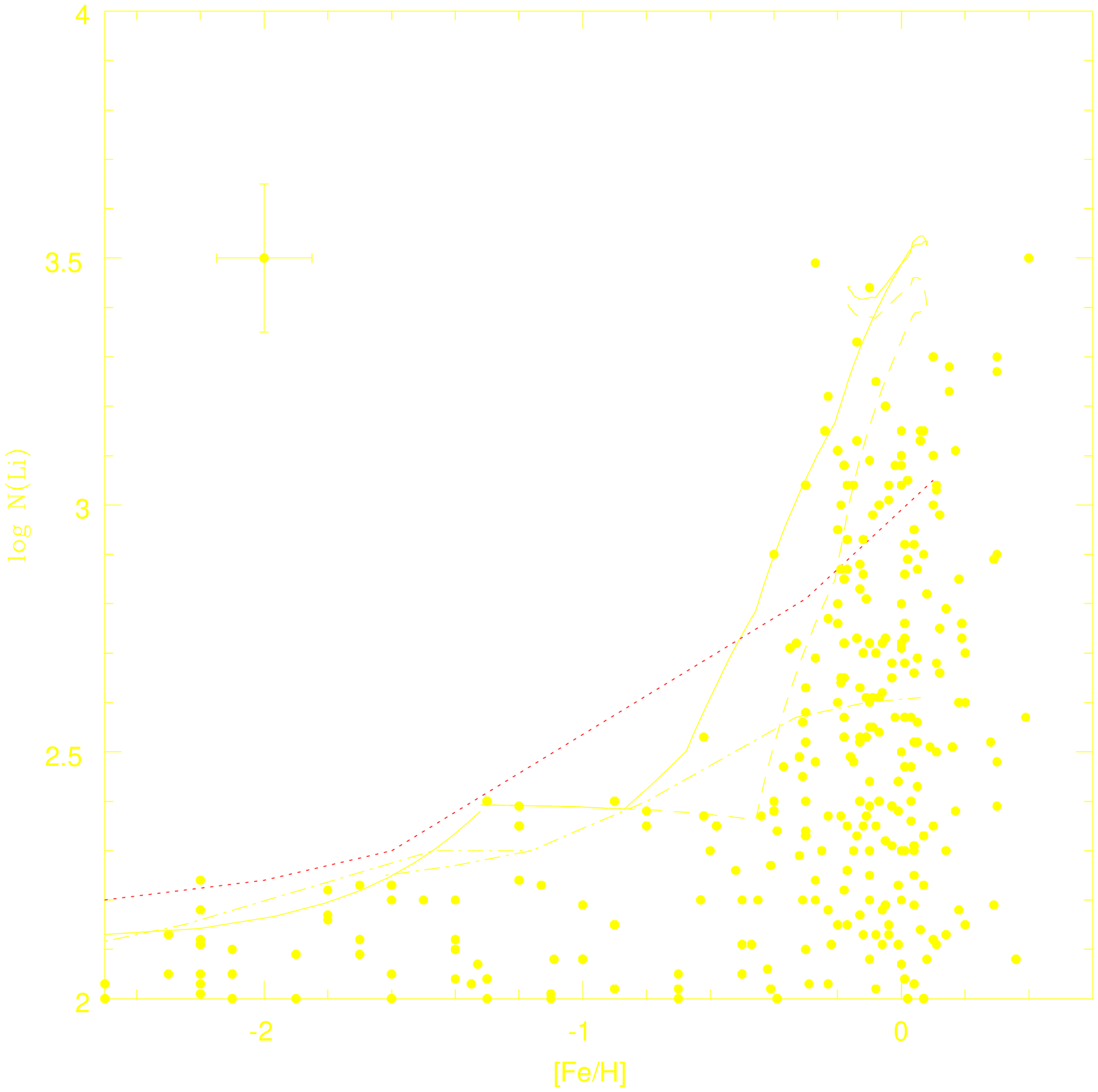}{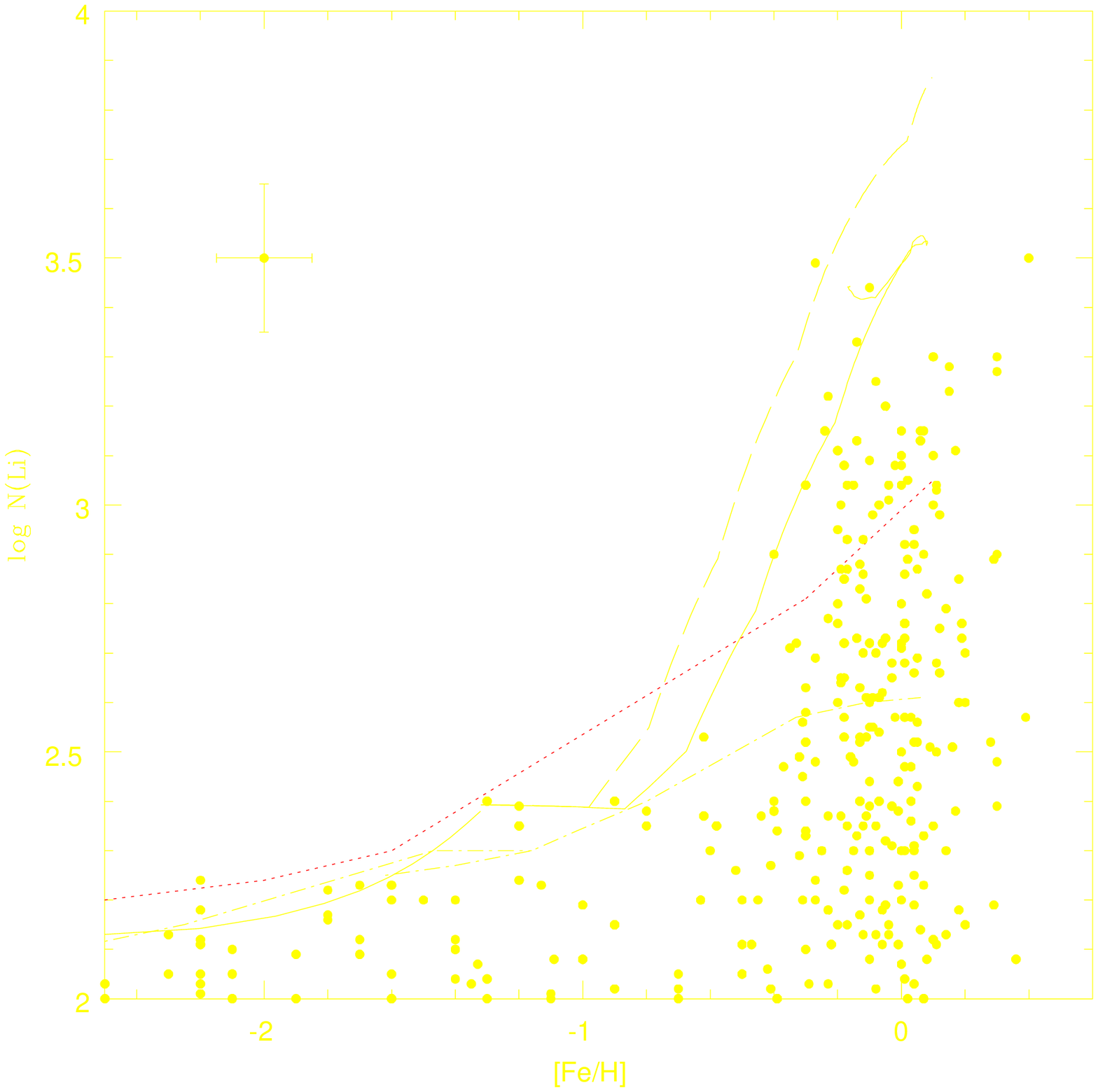}
\label (b)
\newline
\label (c)
\plottwo{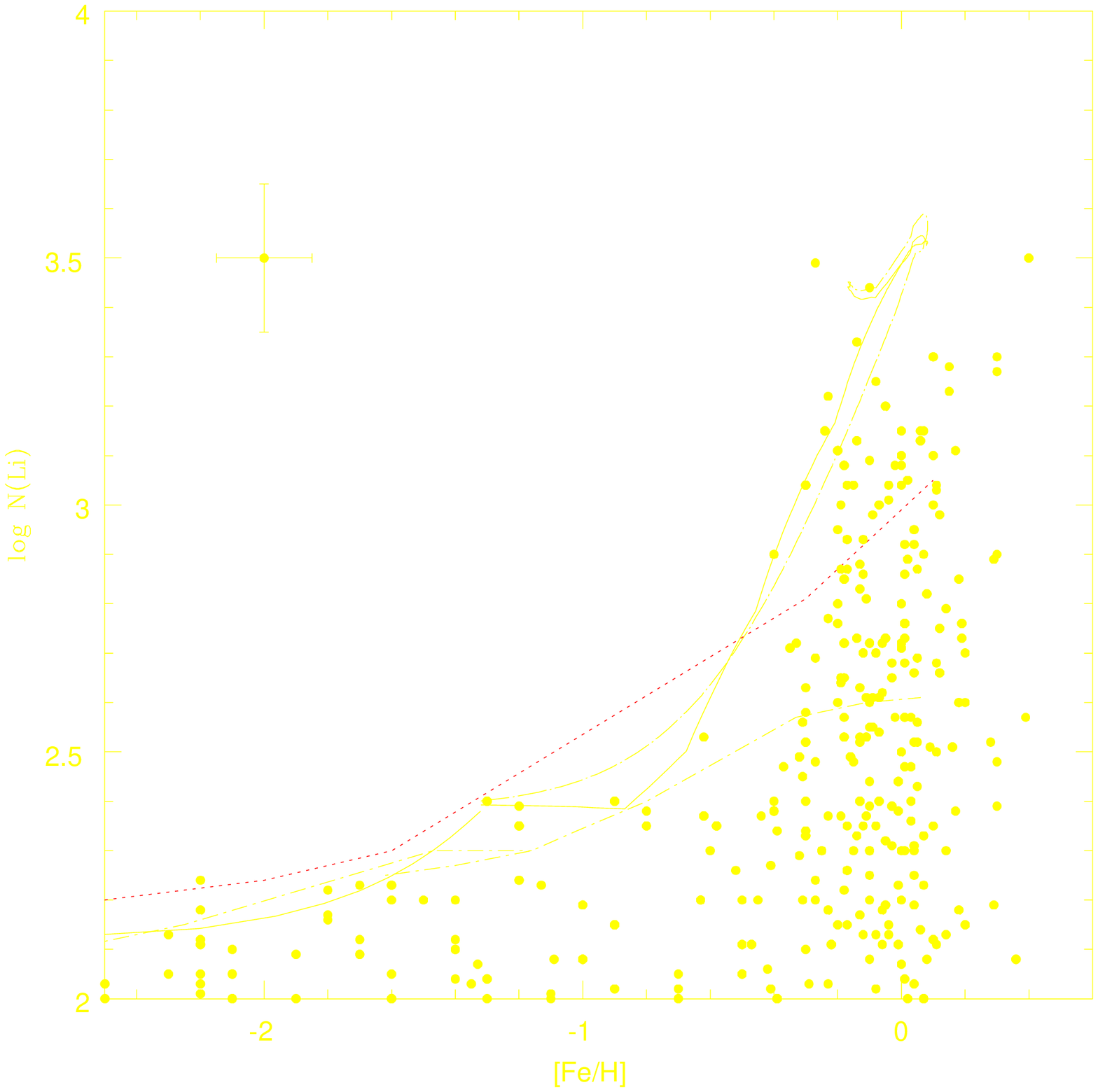}{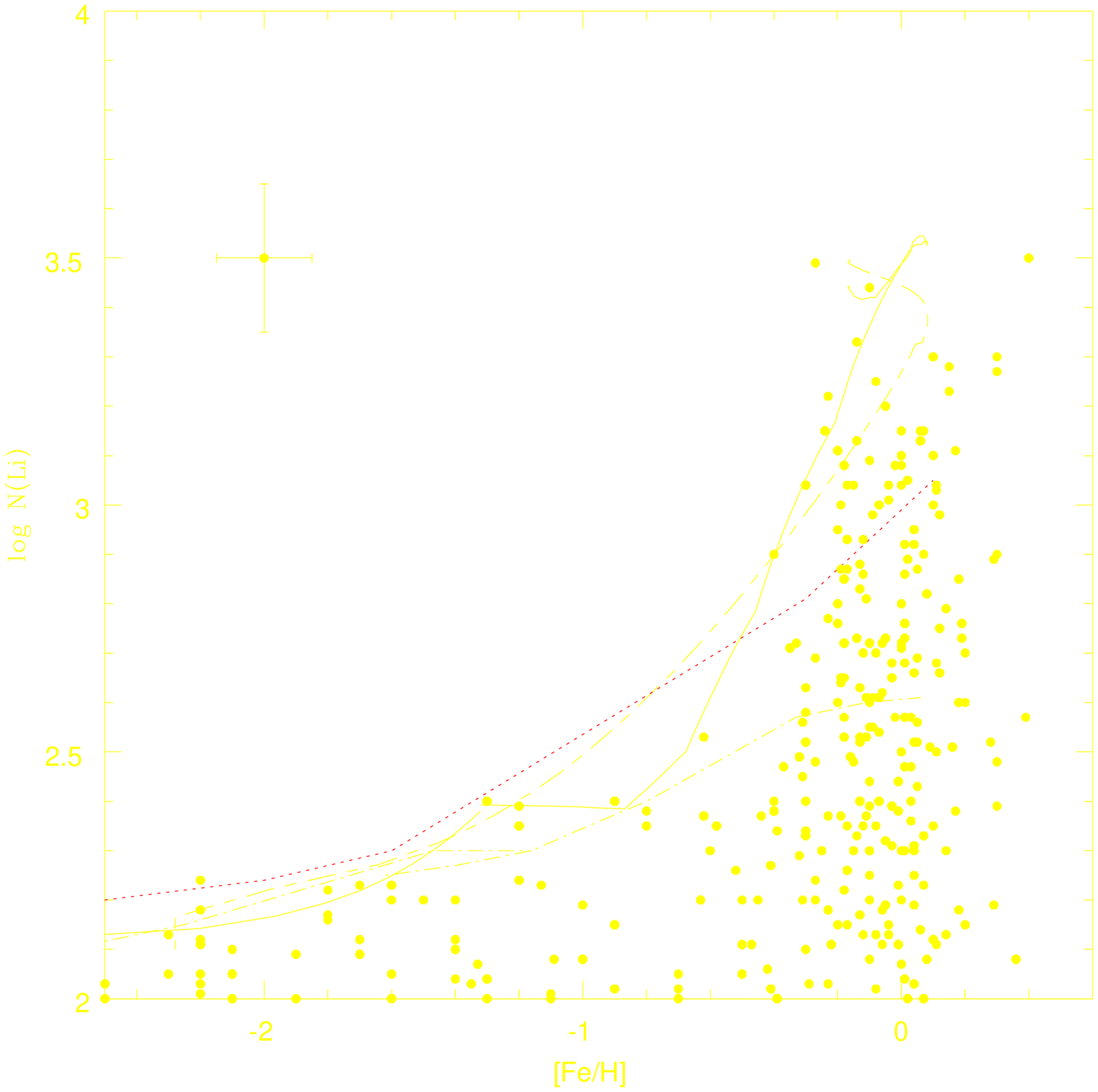}
\label (d)
\end{figure}
\begin{figure}
\plotone{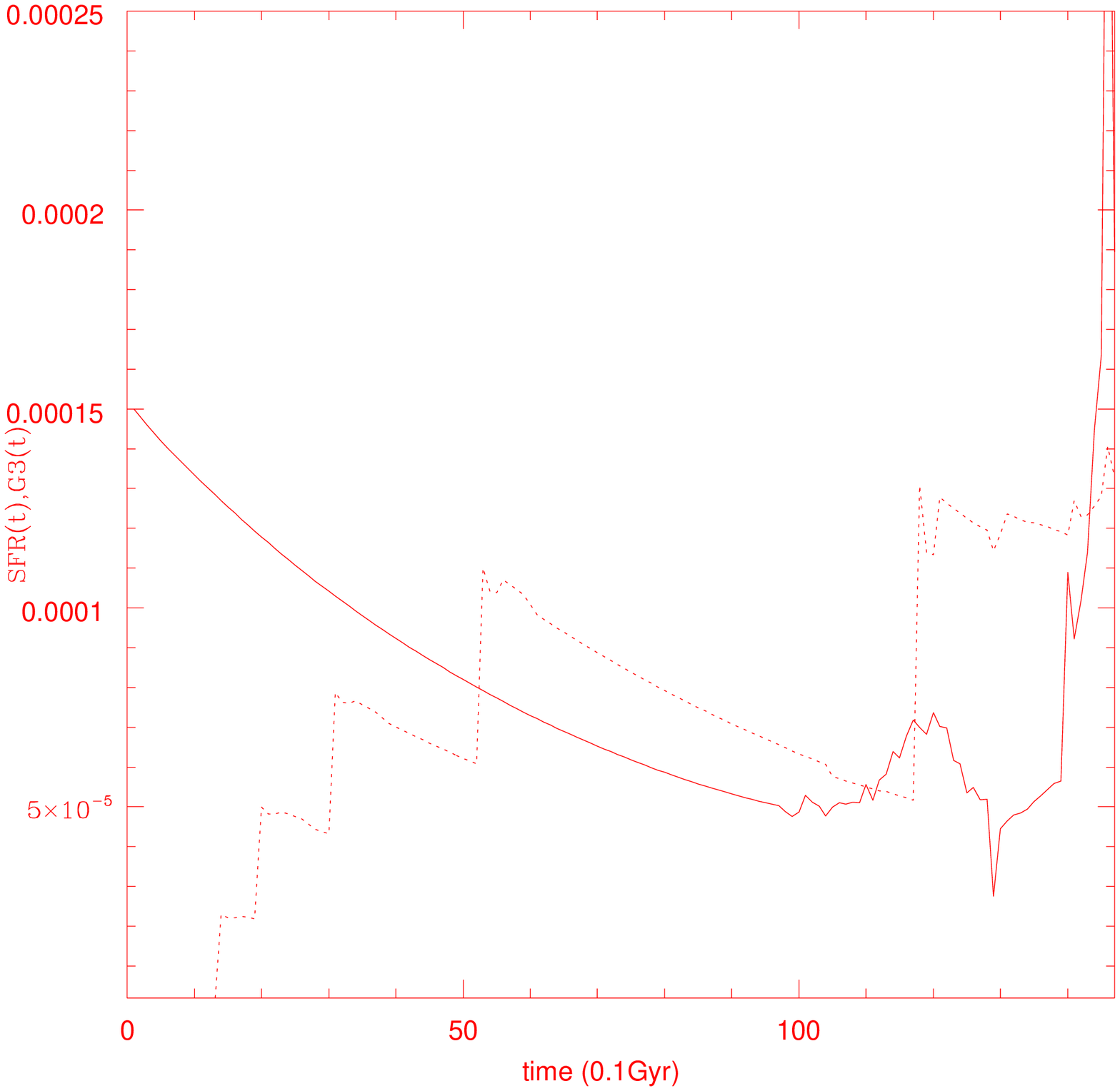}
\end{figure}
\begin{figure}
\plotone{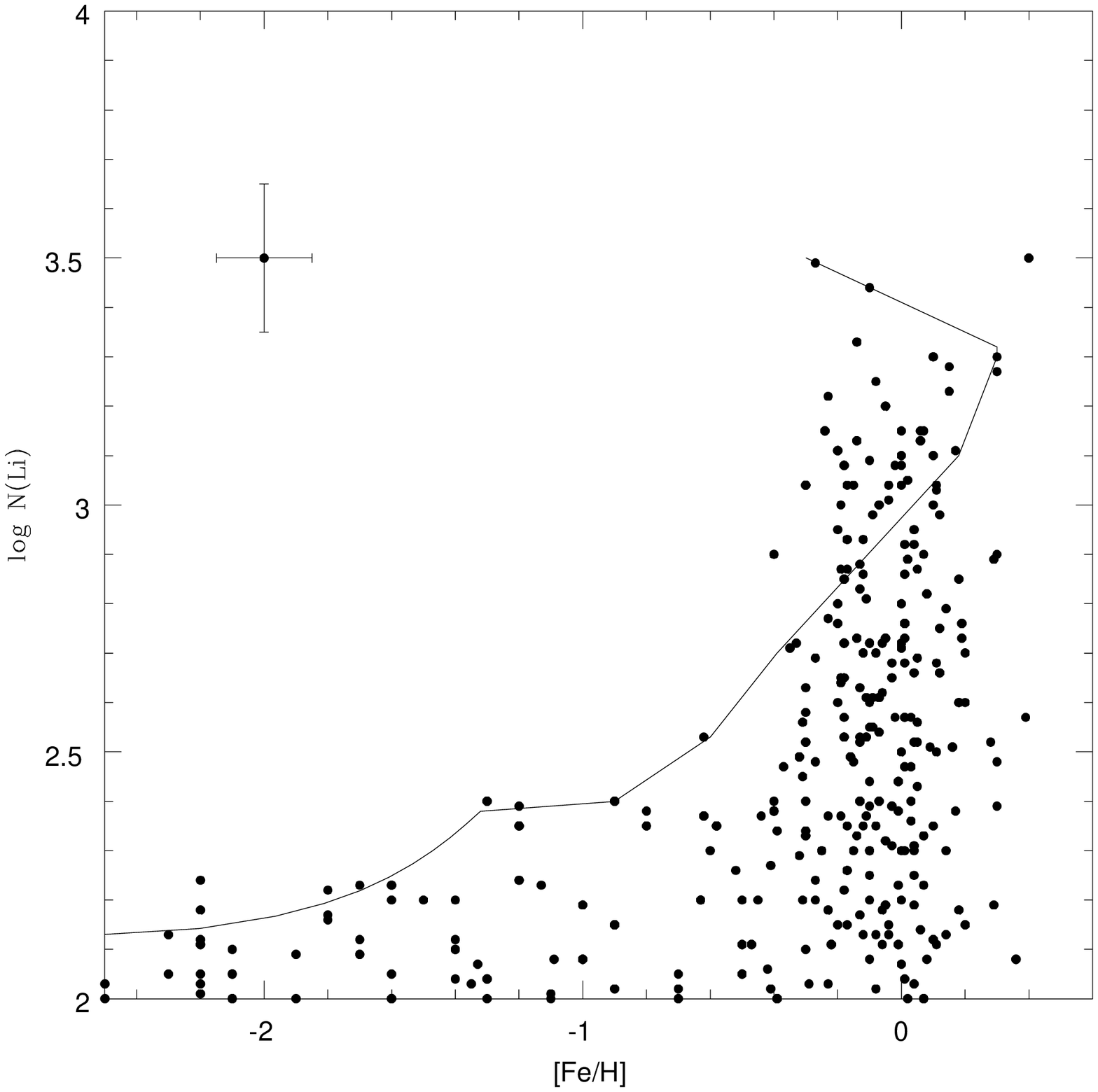}
\end{figure}

\end{document}